\documentclass[twocolumn,prl,amsmath,amssymb,superscriptaddress]{revtex4-1}
\usepackage{graphicx,dsfont,color}
\usepackage{bm}
\usepackage[colorlinks=true,citecolor=blue,urlcolor=blue]{hyperref}
\DeclareMathOperator{\arccot}{arccot}

\begin{document}

\title{Bandgap-Assisted Quantum Control of Topological Edge States in a Cavity}

\author{Wei Nie}
\affiliation{Institute of Microelectronics, Tsinghua University, Beijing 100084, China}
\author{Yu-xi Liu}\email{yuxiliu@mail.tsinghua.edu.cn}
\affiliation{Institute of Microelectronics, Tsinghua University, Beijing 100084, China}
\affiliation{Frontier Science Center for Quantum Information, Beijing, China}

\begin{abstract}
Quantum matter with exotic topological order has potential applications in quantum computation. However, in present experiments, the manipulations on topological states are still challenging. We here propose an architecture for optical control of topological matter. We consider a topological superconducting qubit array with Su-Schrieffer-Heeger (SSH) Hamiltonian which couples to a microwave cavity. Based on parity properties of the topological qubit array, we propose an optical spectroscopy method to observe topological phase transition, i.e., edge-to-bulk transition. This new method can be achieved by designing cavity-qubit couplings. A main purpose of this work is to understand how topological phase transition affects light-matter interaction. We find that topological bandgap plays an essential role on this issue. In topological phase, the resonant vacuum Rabi splitting of degenerate edge states coupling to the cavity field is protected from those of bulk states by the bandgap. In dispersive regime, the cavity induced coupling between edge states is dominant over couplings between edge and bulk states, due to the topological bandgap. As a result, quantum interference between topological edge states occures and enables single-photon transport through boundaries of the topological qubit array. Our work may pave a way for topological quantum state engineering.
\end{abstract}

\maketitle

\textit{Introduction.}---Characterization of topological matter is a crucial issue in condensed matter physics~\cite{Bansil2016Colloquium}. A hallmark of topological phases is the existence of topological invariants, e.g., Chern number and Zak phase, defined on energy bands of the systems~\cite{PhysRevLett.49.405,PhysRevLett.62.2747,RevModPhys.82.1959}. According to edge-bulk correspondence, topological states emerge in the bandgaps and give rise to many novel transport phenomena~\cite{PhysRevLett.103.237001,PhysRevLett.104.056402}. Due to their insensitivity to local decoherence, topological states have prospective applications in quantum information processing. In particular, zero-dimensional edge states, e.g., Majorana bound states are candidate to realize topological quantum computation~\cite{PhysRevB.95.235305,PhysRevLett.120.220504,PhysRevA.98.012336}, and have been observed experimentally in a range of materials, including semiconductor nanowires~\cite{Mourik2012,Deng2012,Albrecht2016,Zhang2018}, ferromagnetic atomic chains~\cite{Perge2014} and iron-based superconductors~\cite{wang2018}. However, the manipulations of edge states are rather challenging, for which reason topological materials with large bandgaps are explored~\cite{xia2009observation,PhysRevLett.109.266801,PhysRevLett.111.136804}.

Cavity quantum electrodynamics (QED), in which quantized electromagnetic fields are strongly coupled to an atomic system, was originally used for studying fundamentals of atomic physics and quantum optics~\cite{Haroche2006book}. With the superb control of quantum states, cavity QED is now applied to quantum information processing, in which the cavity field is proposed for manipulating, measuring, or transferring quantum states of atomic systems~\cite{RevModPhys.85.553,RevModPhys.87.1379}. Circuit QED, in which a microwave transmission line resonator acting as a cavity is coupled to superconducting quantum circuit, is an extension of the cavity QED~\cite{PhysRevA.69.062320,Wallraff2004}. The on-chip circuit QED system is not only a good platform for studying fundamental physics in microwave regime~\cite{Gu2017Microwave}, but also a very promising candidate for realizing quantum computation and simulations~\cite{Devoret2013,PhysRevB.76.174519,PhysRevA.79.040303,PhysRevLett.105.167001,PhysRevLett.110.030601,PhysRevLett.111.110504,2014You,PhysRevLett.112.180405,PhysRevA.98.042328,Macha2014,Barends2016,PhysRevLett.117.210503,
PhysRevX.7.011016,roushan2017spectroscopic,PhysRevLett.120.050507,Mohammad2018,Ma2019,Yan2019}. In particular, one-dimensional (1D) qubit arrays have been used to explore many-body localization~\cite{roushan2017spectroscopic,PhysRevLett.120.050507}, Mott insulator of photons~\cite{Ma2019} and correlated quantum walk~\cite{Yan2019}. Moreover, superconducting qubit systems are also hopeful to simulate topological matter~\cite{PhysRevLett.117.213603,gu2017,PhysRevA.98.012331,PhysRevA.98.032323,nie2019topological}.

In this work, we study the interaction between a microwave cavity and a topological superconducting qubit array, described by the SSH model~\cite{PhysRevLett.42.1698} which has been experimentally realized~\cite{Atala2013,Cai2019,Sylvain2019}. Different from the electronic transport detections of Majorana fermions~\cite{Zhang2018,PhysRevB.86.180503,PhysRevLett.109.267002,PhysRevB.87.024515}, the cavity spectroscopy method we study here unveils the edge states and topological phase transition with proper cavity-qubit couplings. We pinpoint the role of topological bandgap in quantum manipulation of edge states, especially for small qubit arrays.

\textit{Spectroscopic characterization of a topological qubit array by a cavity.}---As schematically shown in Fig.~\ref{figure1}(a), we study that a 1D topological qubit array~\cite{shen2017}, with SSH interactions, is placed inside a cavity. Considering rapid progresses and flexible chip designs of superconducting quantum circuits, we here assume that the SSH array with $N$ unit cells, formed by $2N$ superconducting qubits~\cite{gu2017}, e.g., Xmon qubits~\cite{Barends2016,roushan2017spectroscopic}, is coupled to a microwave transmission line resonator, as shown in Fig.~\ref{figure1}(b). The Hamiltonian of the whole system is
\begin{eqnarray}
H/\hbar =&& \omega_c \hat{a}^{\dagger} \hat{a} + \sum_{i=1,\mu=A,B}^{i=N} \big(\omega_0 \sigma_{i\mu}^+ \sigma_{i\mu}^- + g_{i\mu} \sigma_{i\mu}^+ \hat{a} +   g_{i\mu}^* \hat{a}^{\dagger} \sigma_{i\mu}^- \big) \nonumber \\
&&+\sum_{i=1}^{i=N} \big(t_{1} \sigma_{iA}^+ \sigma_{iB}^- + t_2 \sigma_{i+1A}^+ \sigma_{iB}^- + \mathrm{H.c.} \big), \label{H1}
\end{eqnarray}
where $\omega_c$ and $\omega_0$ are the frequencies of the cavity and qubits, respectively. The parameter $g_{i\mu}$ denotes the coupling strength of the cavity to the qubit $\mu$ in the $i$th unit cell. The operators of qubits $A$ and $B$ at the $i$th unit cell are $\sigma_{iA}^+ = |A_i\rangle \langle \alpha_i |$ and $\sigma_{iB}^+ = |B_i\rangle \langle \beta_i |$ with  the ground (excited) states  $|\alpha_i\rangle$ ($|A_i\rangle$) and $|\beta_i\rangle$ ($|B_i\rangle$), respectively. The second line in Eq.~(\ref{H1}) represents the SSH interaction Hamiltonian with tunable coupling strengths $t_1$ and $t_2$, which could be implemented with different ways in superconducting qubit circuits~\cite{PhysRevLett.90.127901,PhysRevLett.96.067003,PhysRevB.74.172505,PhysRevLett.98.057004,Niskanen2007,Majer2007,PhysRevLett.113.220502,roushan2017spectroscopic,PhysRevLett.120.050507,PhysRevA.92.012320}. We here assume that controllable coupling between qubits is realized via a Josephson junction, which is biased by an external magnetic flux~\cite{PhysRevLett.113.220502,PhysRevA.92.012320}, as shown in Fig.~\ref{figure1}(c). The  coupling strengths are $t_1=t_0(1-\cos \varphi)$ and $t_2=t_0(1+\cos\varphi)$ with a tunable parameter $\varphi$~\cite{SupplementalMaterial}. Note that the topological phase transition takes place at $\varphi=\pi/2$ ($t_1=t_2$). The cases for $t_1<t_2$ and $t_1>t_2$ correspond to topological and non-topological phases, respectively.

\begin{figure}[t]
\includegraphics[width=8.7cm]{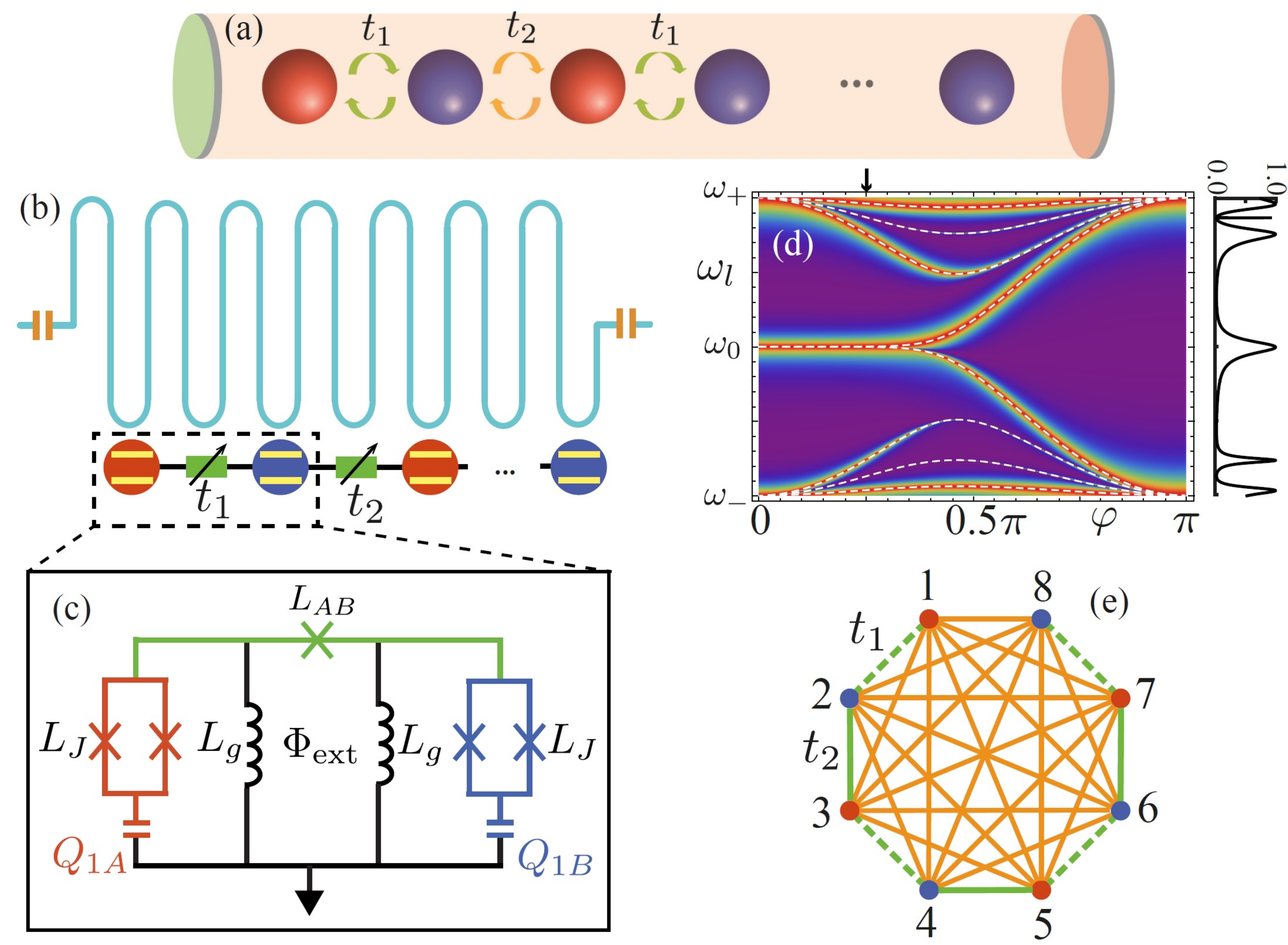}
\caption{(a) Schematic diagram for SSH qubit array with dimerized couplings $t_{1}$ and $t_{2}$  placed inside a cavity. Red and blue balls represent $A$ and $B$ qubits, respectively. (b) Design of (a) with superconducting qubit circuits where the couplings $t_1$ and $t_2$ are tunable. The microwave transmission line resonator acting as a cavity is coupled to qubits. (c) The first unit cell with tunable qubit-qubit coupling via a Josephson junction. $\Phi_{\mathrm{ext}}$ is an external magnetic flux bias that controls the coupling (see Supplemental Material~\cite{SupplementalMaterial}). (d) Reflection spectrum of the qubit array with $8$ qubits. The frequencies of qubits and driving field are respective $\omega_0$ and $\omega_l$; $\omega_{\pm}=\omega_0 \pm 2t_0$. The reflection at $\varphi=0.25\pi$ is shown in the right panel. Here we consider cavity-qubit couplings $\bm{g}=g_0(-1,1,1,1,-1,1,1,1)$ with $g_0/2\pi=5$ MHz. Other parameters are: $\omega_0/2\pi=6$ GHz, $t_0/2\pi= 100$ MHz, $\kappa/2\pi=10$ MHz, $\gamma_{iA}=\gamma_{iB}=20 \times 2\pi$ kHz. The white-dashed curves represent energy spectrum of the qubit array. (e) Cavity mediated couplings between qubits, denoted by the orange lines,  in dispersive regime.}\label{figure1}
\end{figure}

To measure the topological qubit array, we assume that a weak probe field with the strength $\eta$ and the frequency $\omega_{l}=\omega_{c}$ is applied to the qubit array via the cavity. Thus, the dynamics of the reduced density matrix $\rho$ of the whole system can be described by the master equation
\begin{eqnarray}\label{eq:2}
\dot{\rho} = &&-\frac{i}{\hbar}\Big[H+i \hbar\eta(\hat{a}^{\dagger}e^{-i\omega_l t} - \hat{a}e^{i \omega_l t}), \rho\Big]  \nonumber \\
&&+ \kappa \mathcal{D}[\hat{a}]\rho + \sum_{i=1,\mu=A,B}^{i=N}\gamma_{i\mu} \mathcal{D}[\sigma_{i\mu}^-]\rho.
\end{eqnarray}
\begin{figure}[b]
\includegraphics[width=8.5cm]{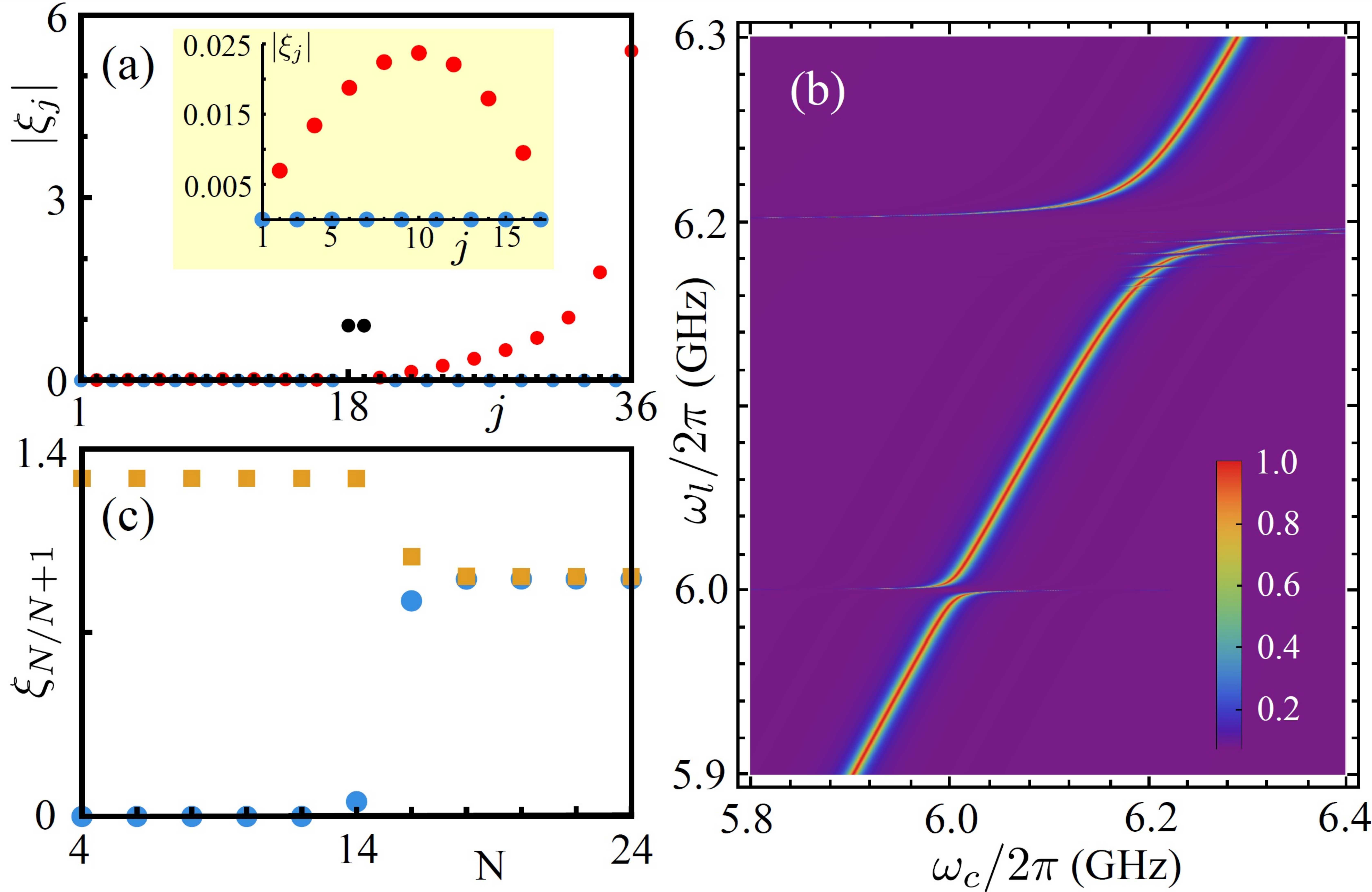}
\caption{(a) The coupling strength between the cavity and eigenmodes of SSH qubit array with $N=18$ unit cells (i.e. 36 qubits) for homogeneous cavity-qubit couplings. The number $j$ is the index of eigenmodes, and the middle two points (black dots) with $j=18,19$ are edge states. The bulk states with odd (blue dots) and even (red dots) numbers have zero and nonzero couplings to the cavity, respectively. The inset represents the coupling strengths for $j<18$.  (b) Vacuum Rabi splitting between edge states and the cavity for $N=18$ unit cells. The lower anticrossing shows the coupling between the cavity and the edge states. The upper one is the coupling of the bulk states to the cavity. (c) The coupling strengths $
\xi_{N}$ and $\xi_{N+1}$, denoted by $\xi_{N/N+1}$, between two edge modes and the cavity with different numbers $N$ of unit cells. Here we consider $\varphi=\pi/5$. Other parameters are the same to Figure~\ref{figure1}(d). }\label{figure2}
\end{figure}
Here, $\kappa$ is the decay rate of the cavity, $\gamma_{iA}$ and $\gamma_{iB}$ are the decay rates of the qubits $A$ and $B$ at the $i$th unit cell, respectively. The dissipation superoperator is defined as $\mathcal{D}[\hat{O}]\rho = \hat{O}\rho \hat{O}^{\dagger} -\frac{1}{2}\{\hat{O}^{\dagger}\hat{O}, \rho\}$. As shown in Fig.~\ref{figure1}(d), the topological phase transition can be observed from the reflection of the probe field with special couplings between qubits and cavity, which can be realized via controllable couplers~\cite{SupplementalMaterial,Zhong2019NP,Li2019CPB}. The reflection spectrum is obtained by solving the master equation in Eq.~(\ref{eq:2}) with $H$ given in Eq.~(\ref{H1}). Topological phases have recently been demonstrated in superconducting qubit circuits~\cite{PhysRevLett.113.050402,Roushan2014,PhysRevX.7.031023,Wang2018Simulating,PhysRevLett.120.130503,PhysRevLett.121.030502,King2018Observation,PhysRevLett.122.010501,Cai2019}. However, the operations on topological states have not been implemented. Below we show that topological bandgap is helpful for manipulations of edge states.

\textit{Vacuum Rabi splitting between the cavity and edge modes}.---We consider the single-excitation subspace consisted of $|\mathcal{A}_i\rangle= \sigma_{iA}^+ |G\rangle$ and $|\mathcal{B}_i\rangle= \sigma_{iB}^+ |G\rangle$ with $|G\rangle$ being the ground state of the qubit array. We rewrite the states $|\mathcal{A}_i\rangle$ and $|\mathcal{B}_i\rangle$ via eigenstates $|\Psi_{j}\rangle$ in the single-excitation subspace of the qubit array~\cite{nie2019topological}, i.e., $|\mathcal{A}_i\rangle=\sum_{j=1}^{2N} \xi_{2i-1, j} |\Psi_j\rangle$ and $|\mathcal{B}_i\rangle=\sum_{j=1}^{2N} \xi_{2i, j} |\Psi_j\rangle$. Here, $j=1,\cdots, 2N$ is the label of the $j$th eigenstate from the lowest to highest energies. Then, the Hamiltonian in Eq.~(\ref{H1}) can be rewritten as~\cite{SupplementalMaterial}
\begin{equation}
\tilde{H}/\hbar = \sum_{j=1}^{2N} \omega_j \Psi_j^+\Psi_j^- + \omega_c \hat{a}^{\dagger}\hat{a}  +  \sum_{j=1}^{2N} (\tilde{\xi}_j \Psi_j^+ \hat{a} + \mathrm{H.c.}),
\end{equation}
with $\Psi_j^+ = |\Psi_j\rangle \langle G|$, and the eigenfrequency $\omega_j$ corresponding to the eigenstate $ |\Psi_j\rangle$. For homogeneous cavity-qubit couplings, i.e., $g_{i\mu}\equiv g_0$, the coupling strength between cavity and the $j$th eigenmode is $\tilde{\xi}_j=\xi_j g_0$ with coupling coefficient $\xi_j=\sum_i (\xi_{2i-1,j} + \xi_{2i,j})$. The analytical expressions for $\xi_j$ can be found in Ref.~\cite{SupplementalMaterial}. Hereafter, we call $\Psi_j^+ $ bulk or edge modes when $|\Psi_j\rangle$ are bulk or edge eigenstates. In Fig.~\ref{figure2}(a), we show $|\xi_j|$ for the  qubit array size $2N=36$. The bulk modes have different couplings to the cavity because of their parities of wavefunctions. The odd-parity bulk states, i.e., $\xi_{2i-1,j}=-\xi_{2N+2-2i,j}$ and $\xi_{2i,j}=-\xi_{2N+1-2i,j}$, have zero coupling to the cavity. However, the even-parity bulk states, i.e., $\xi_{2i-1,j}=\xi_{2N+2-2i,j}$ and $\xi_{2i,j}=\xi_{2N+1-2i,j}$, are coupled to the cavity~\cite{SupplementalMaterial}. Two edge states have equal coupling strength to the cavity, i.e., $\xi_{18}=\xi_{19}$.

\begin{figure}[b]
\includegraphics[width=8cm]{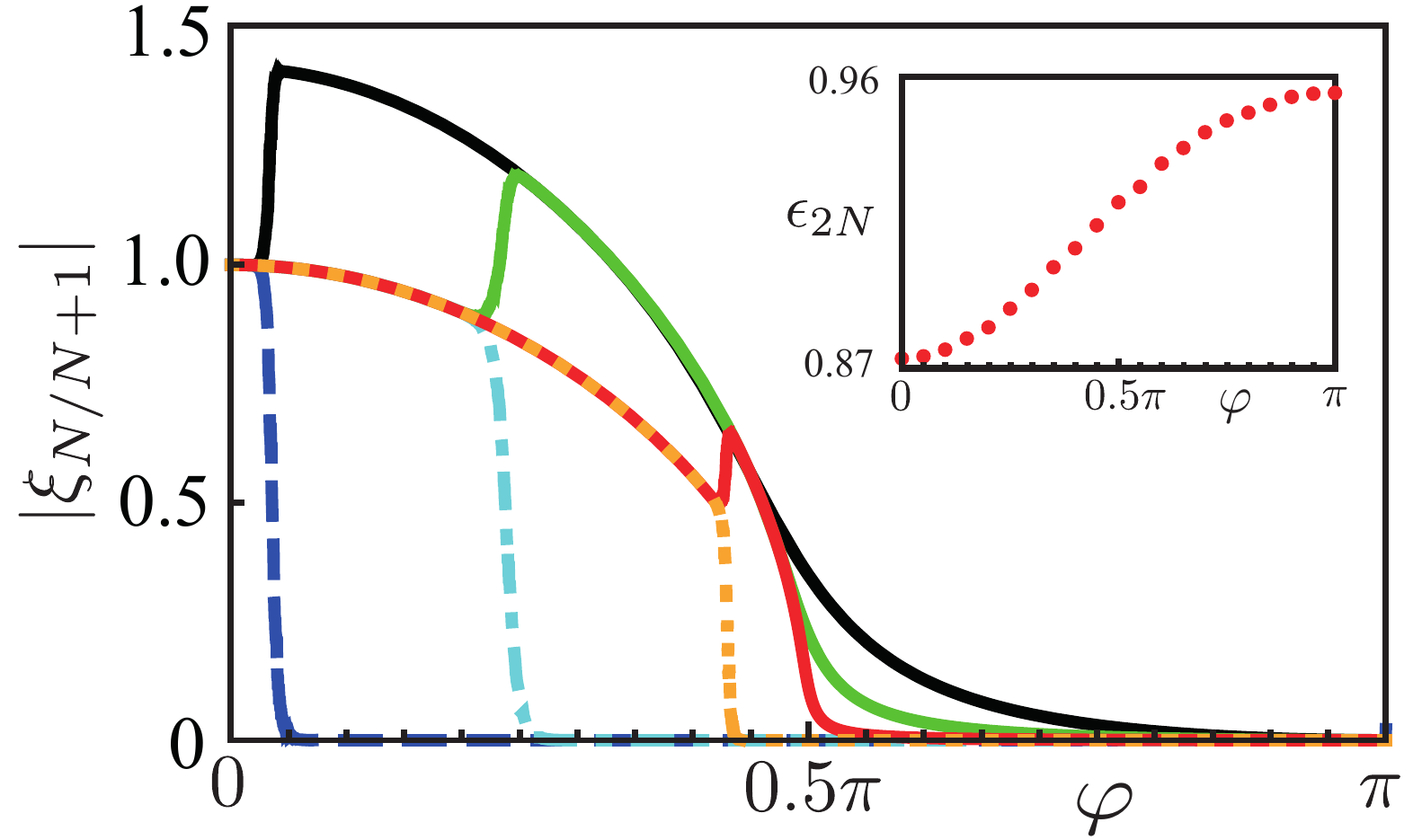}
\caption{The couplings between the cavity and edge states (black-solid and blue-dashed) for the qubit array with $N=6$ unit cells. We also present the coupling strengths for $N=18$ ($N=78$) by the green-solid and blue-dash-dotted (red-solid and orange-dotted) curves. The inset shows the coefficient $\epsilon_{2N}$ of the collective coupling $\tilde{\xi}_{j}=\epsilon_{j}\sqrt{2N}g_0$ with $j=2N$.}\label{figure3}
\end{figure}

In Fig.~\ref{figure2}(b), we show energy splitting produced by the cavity-qubit couplings. We assume that the qubit frequency is  $\omega_0 =2\pi\times6$ GHz. The anticrossing near the driving frequency $\omega_l=2\pi\times6$ GHz represents the Rabi splitting due to the resonant interaction between the cavity and edge modes. We also study the disorder effect and find that the Rabi splitting of edge states is robust to disorder~\cite{SupplementalMaterial}. If the frequency of the cavity is at resonance for the transitions from the ground to bulk states with high energies, a large anticrossing,  as shown in upper part of Fig.~\ref{figure2}(b), is produced around $\omega_l=2\pi\times6.2$ GHz. The topological bandgap of SSH Hamiltonian protects the Rabi splitting of edge states. In Fig.~\ref{figure2}(c), $\xi_{N}$ and $\xi_{N+1}$, i.e., the coupling coefficients between the cavity and edge modes, are plotted versus the unit cell number $N$. When the qubit array is small, e.g., $N\leq14$, the edge states overlap with each other and form hybridized edge states with odd and even parities. The edge state with odd parity decouples from the cavity. With the increase of the unit cell number, two edge states are far separated from each other. The localized edge states lose parity, thus they have the same coupling strength to the cavity.

\begin{figure}[t]
\includegraphics[width=8.5cm]{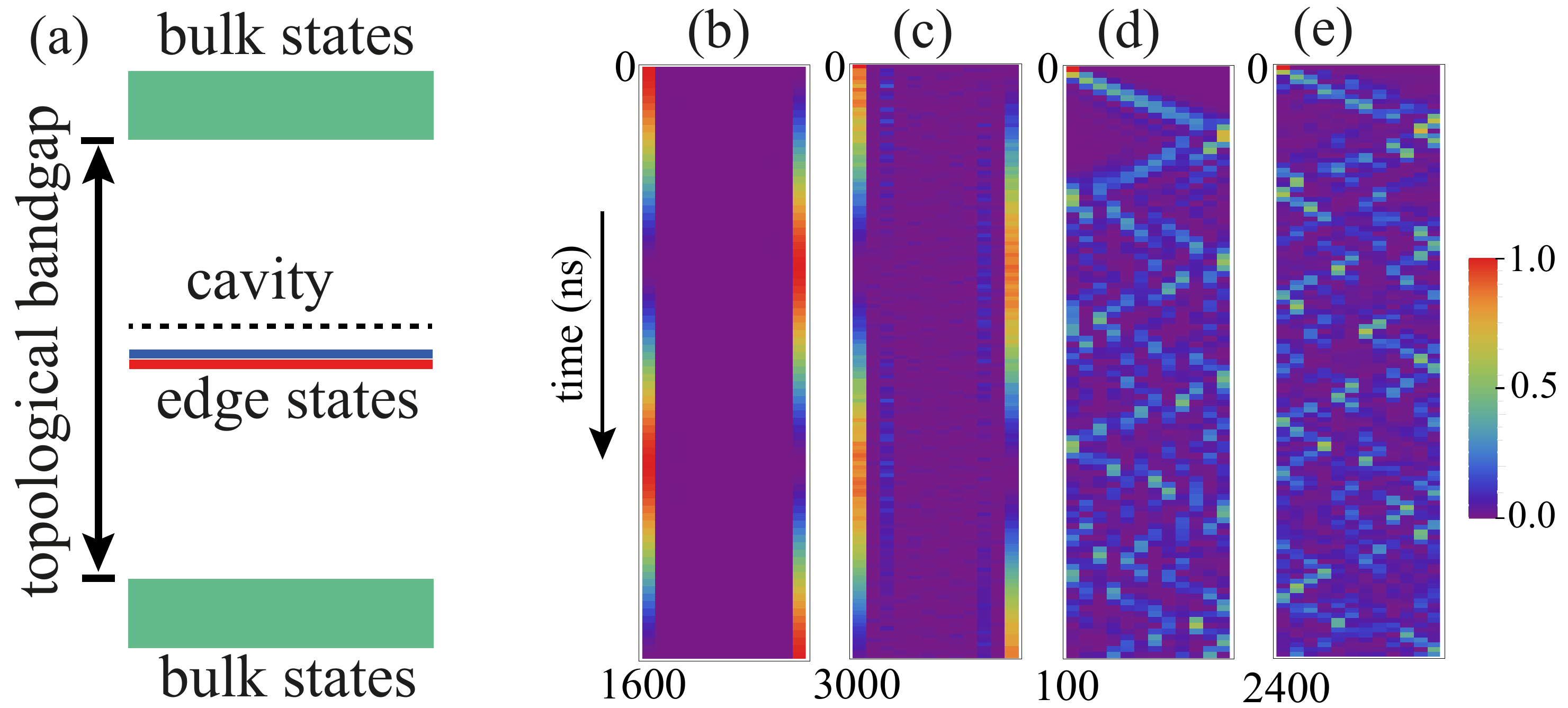}
\caption{(a) Edge-state coupling protected by topological bandgap. (b)-(e) Excitation dynamics in the cavity-coupled SSH qubit array for $\varphi=0.1\pi, 0.3\pi,0.5\pi$ and $0.9\pi$. Horizontal and vertical axes denote qubits and time, respectively. The virtual-photons-mediated interactions among qubits are assumed to be $0.1 g_0$. The number of unit cells is $N=6$, and other parameters are as the same as Figure~\ref{figure1}(d).}\label{figure4}
\end{figure}

We study the relation between the coupling coefficient $\xi_{N}$ ($\xi_{N+1}$) and $\varphi$ in Fig.~\ref{figure3}. For example, when the qubit array has $N=6$ unit cells, the coupling strengths are described by the black-solid and blue-dashed curves. When $\varphi$ is small, the edge states are unhybridized and have the same coupling to the cavity. However, the increase of $\varphi$ leads to hybridized edge states with even and odd parities. We find that the hybridized regime becomes smaller with the increase of the system size, e.g., $N=18$ (green-solid and blue-dash-dotted curves) and $N=78$ (red-solid and orange-dotted curves) as we show here. We also find that in topological phase (i.e., $0\leq\varphi<\pi/2$), the hybridized edge state with even parity has the coupling strength $\tilde{\xi}_+=\sqrt{2\cos\varphi} g_0$, which is independent of system size~\cite{SupplementalMaterial}. The couplings for separated edge states are $\tilde{\xi}_L=\tilde{\xi}_R=\sqrt{\cos\varphi} g_0$. Because the coupling strength of $2N$ non-interacting qubits to the cavity is $\sqrt{2N}g_0$~\cite{PhysRev.170.379}, we here assume that the coupling strength of $2N$ interacting qubits to the cavity is $\tilde{\xi}_{j}=\epsilon_{j}\sqrt{2N}g_0$ with the rescaling factor $\epsilon_{j}$ given in Eq.~(S29)~\cite{SupplementalMaterial}. As shown in the inset of Fig.~\ref{figure3}, the rescaling factor $\epsilon_{2N}$ is tuned by $\varphi$.

\textit{Topological-bandgap-protected coupling between two edge modes.}---When the cavity is far detuned from qubits, i.e., $g_0\ll \Delta_0$ (let $\Delta_0=\omega_0-\omega_c$), virtual-photons-mediated interactions among qubits $g_0^2/\Delta_0$ can be obtained~\cite{Majer2007,PhysRevLett.120.050507}, as shown in Fig.~\ref{figure1}(e). In terms of eigenmodes of the qubit array, the effective coupling strengths between $j$th and $k$th eigenmodes are
\begin{equation}
J_{jk}=\frac{\tilde{\xi}_j \tilde{\xi}_k}{2}\Big( \frac{1}{\Delta_j} + \frac{1}{\Delta_k} \Big), \quad j,k \in [1,\cdots, 2N],
\end{equation}
with $\Delta_{j/k}=\omega_{j/k}-\omega_{c}$, which depends on system size $2N$ for the coupling between bulk modes, or between bulk and edge modes, due to the size-dependent cavity-bulk coupling. However, the edge-mode coupling is independent of the size, i.e.,
\begin{equation}\label{Eqedge}
J=\cos\varphi\frac{g_0^2}{\Delta_0},
\end{equation}
which is protected by the topological bandgap, as schematically shown in Fig.~\ref{figure4}(a). We note that the energy splitting induced by hybridization of edge states is assumed to be negligibly small when Eq.~(\ref{Eqedge}) is derived. In Figs.~\ref{figure4}(b) and \ref{figure4}(c), we show the excitation dynamics of the left-edge qubit (qubit $A$ in the first unit cell is excited initially) in topological phase with $\varphi=0.1\pi$ and $0.3\pi$, respectively.  Figures~\ref{figure4}(b) and \ref{figure4}(c) clearly show the population exchange between two edge states produced  by the edge-mode coupling. In fact, finite topological bandgap makes the effective couplings between edge modes different from Eq.~(\ref{Eqedge})~\cite{SupplementalMaterial}. In Figs.~\ref{figure4}(d) and \ref{figure4}(e) with $\varphi=0.5\pi$ and $0.9\pi$, the excitation propagates through the array and is bounded by the boundaries. In non-topological phase, excitation propagates along the qubit array with low velocity (see Fig.~\ref{figure4}(e)), which is yielded by the smooth energy bands with large gap.

\textit{Quantum interference induced by topological state coupling.}---As schematically shown in Fig.~\ref{figure5}(a), we further consider that the left-edge qubit $A_1$ is coupled to a waveguide, in which a probe field passes through. The left-edge qubit mainly contributes to the left edge state. Then the left edge state can be driven by fields passing through the waveguide.  The single photons transmission amplitude can be given as~\cite{SupplementalMaterial}
\begin{equation}
t=\frac{(i\Delta_{p}-\frac{\gamma_{L}}{2})  (i\Delta_{p}-\frac{\gamma_{R}}{2}) + J^2}{(i\Delta_{p}-\frac{\gamma_L +\Gamma_{L}}{2})  (i\Delta_{p}-\frac{\gamma_{R}}{2}) + J^2},
\end{equation}
and the susceptibility $\chi=-i(t-1)/t$ is
\begin{equation}
\chi = \frac{\Gamma_L  (\Delta_{p}+ i\frac{\gamma_{R}}{2})}{2J^2-2(\Delta_{p}+i\frac{\gamma_L}{2})  (\Delta_{p}+i\frac{\gamma_{R}}{2})},
\end{equation}
where $\Delta_p$ is the detuning between the probe field and the left edge state. As schematically shown in  Fig.~\ref{figure5}(b), the parameters $\gamma_L$ and $\gamma_R$ are the decay rates for left and right edge states,  $\Gamma_L$ comes from the coupling between the left-edge qubit and the waveguide.

\begin{figure}[t]
\includegraphics[width=8.5cm]{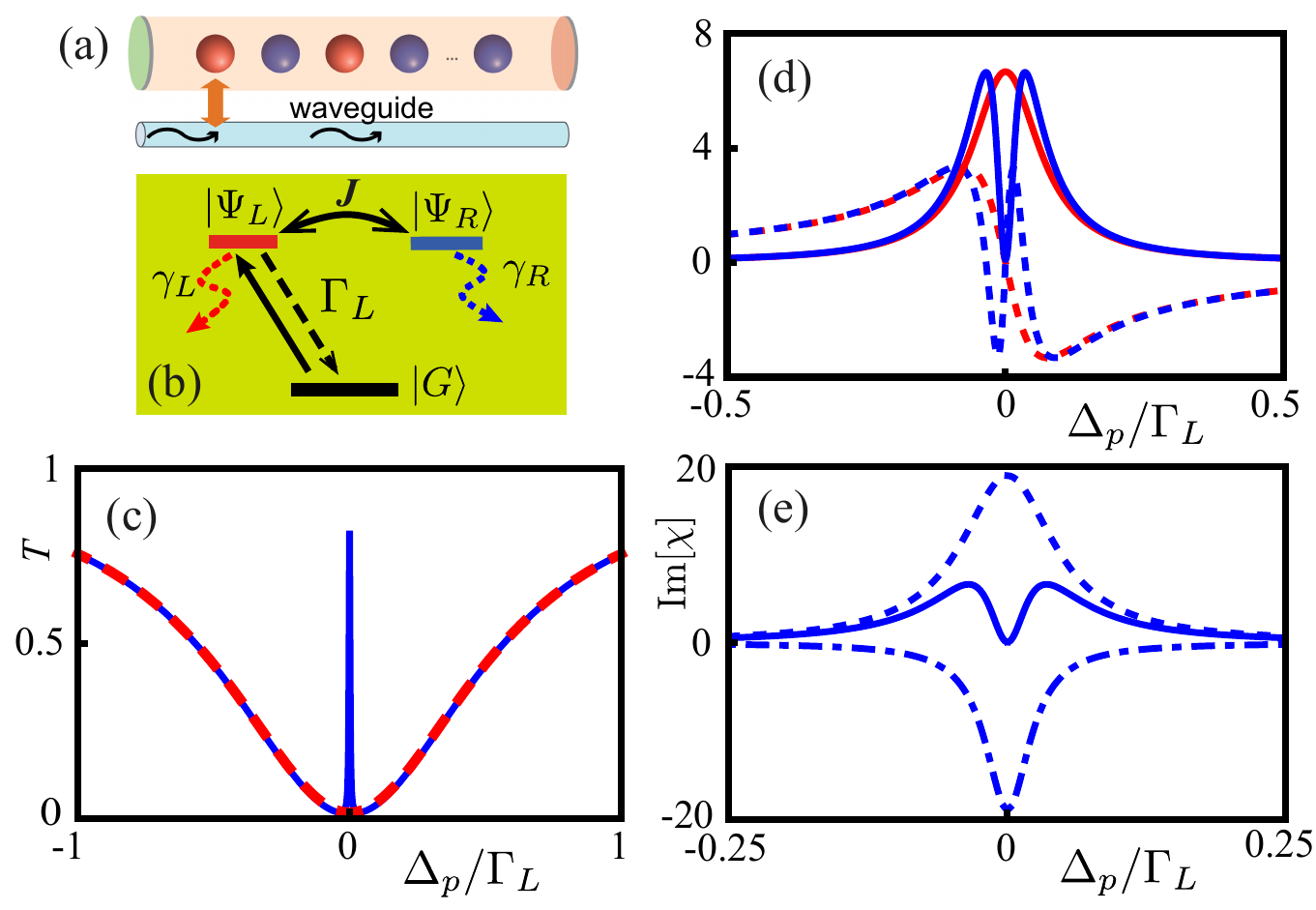}
\caption{(a) Coupling between a waveguide and the left-edge qubit in the array. (b) Superatom model of (a). Here $\gamma_L$ and $\gamma_R$ are respective decays of left and right edge states, and $\Gamma_L$ is the waveguide induced decay rate of the  left edge state. (c) Transmission of probe light for edge-state coupling $J=0$ ($J=0.035 \Gamma_L$) is represented by the red-dashed (blue-solid) curve. (d) Real (solid) and imaginary (dashed) parts of the susceptibility. In both (c) and (d), the red and blue curves are for $J=0$ and $J=0.035 \Gamma_L$, respectively.  (e) The imaginary part of susceptibility can be decomposed into two Lorentzian peaks. In these three figures, we consider $\gamma_L = 0.15 \Gamma_L$, $\gamma_R =5 \times 10^{-4} \Gamma_L$.}\label{figure5}
\end{figure}

The transmission of the probe field as a function of the detuning $\Delta_{p}$ is shown in Fig.~\ref{figure5}(c) with $J=0$ and $0.035 \Gamma_L$, respectively. When there is no coupling between edge states, the transmission vanishes at the resonance. However, when there is the coupling between two edge states, a transparency windows for the probe field appears.  This can be further confirmed by the susceptibility, which is plotted as a function of the detuning $\Delta_{p}$ in Fig.~\ref{figure5}(d) in the parameter regime $J\ll\Gamma_{L}$. This transparency window, in which the distance between two peaks is less than $2J$, is from the quantum interference as shown in Fig.~\ref{figure5}(e), which is similar to electromagnetically induced transparency~\cite{PhysRevLett.120.083602}. However, in the parameter regime $J>\Gamma_{L}$, the transparency window, in which the distance between two peaks equals to $2J$, is from the strong-coupling-induced energy splitting, which is similar to Autler-Townes splitting~\cite{PhysRevLett.103.193601}.

\textit{Discussions and conclusions.}---In summary, we study cavity control of topological edge states in SSH qubit arrays. We show that the coupling between cavity and edge modes are protected by topological bandgap, and topological phase transitions can be probed via the reflection spectrum of the probe field through the cavity. Due to the bandgap, Rabi splitting of edge modes can be observed. When the cavity is largely detuned from the edge modes, long-range coupling between two edge states can be realized. It results in quantum interference for emissions from two edge states when a qubit at the edge of the array is coupled to a waveguide. Meanwhile, we find that topological properties can also be detected by the cavity even for a small system, where the edge states are hybridized.

We also discuss experimental feasibility via superconducting qubit array coupled to microwave transmission line resonator~\cite{gu2017}. The tunable couplings between qubits~\cite{PhysRevLett.113.220502} or between qubits and the resonator~\cite{Zhong2019NP} make our proposal more experimentally accessible. We also analyze the effects of disorder and decay on the results. We find that both the Rabi splitting and excitation dynamics produced by edge-state coupling are robust to the disorder~\cite{SupplementalMaterial}. We also find that with current coherence time $20 \sim 40$ $\mu$s in 1D arrays~\cite{Ma2019,Yan2019}, the Rabi splitting, excitation dynamics, and quantum interference induced by the edge-state coupling  should be observed. We mention that our approach can also be applied to other topological quantum systems. Our study here might have potential applications in quantum information and quantum optics.

\textit{Acknowledgments---.}The authors thank Prof. Xuedong Hu for helpful discussions. Y.X.L. is supported by the Key-Area Research and Development Program of GuangDong Province under Grant No. 2018B030326001, the National Basic Research Program (973) of China under Grant No. 2017YFA0304304, and NSFC under Grant No. 11874037. W.N. acknowledges the Tsinghua University Postdoctoral Support Program.

\clearpage \widetext
\begin{center}
	\section{Bandgap-Assisted Quantum Control of Topological Edge States in a Cavity \\
-- Supplemental Material}
\end{center}
\setcounter{equation}{0} \setcounter{figure}{0}
\setcounter{page}{1} \setcounter{secnumdepth}{3}
\renewcommand{\theequation}{S\arabic{equation}}
\renewcommand{\thefigure}{S\arabic{figure}}
\renewcommand{\bibnumfmt}[1]{[S#1]}
\renewcommand{\citenumfont}[1]{S#1}

\makeatletter
\def\@hangfrom@section#1#2#3{\@hangfrom{#1#2#3}}
\makeatother

\maketitle

\section{Circuit QED with a SSH qubit array}
In the main text, we consider a system where a SSH qubit array, which has controllable interactions between qubits, couples to a microwave transmission line resonator. In this section, we show how this topological circuit QED can be realized with state-of-the-art techniques.

\subsection{SSH qubit array with tunable couplings}
\begin{figure}[b]
\includegraphics[width=12cm]{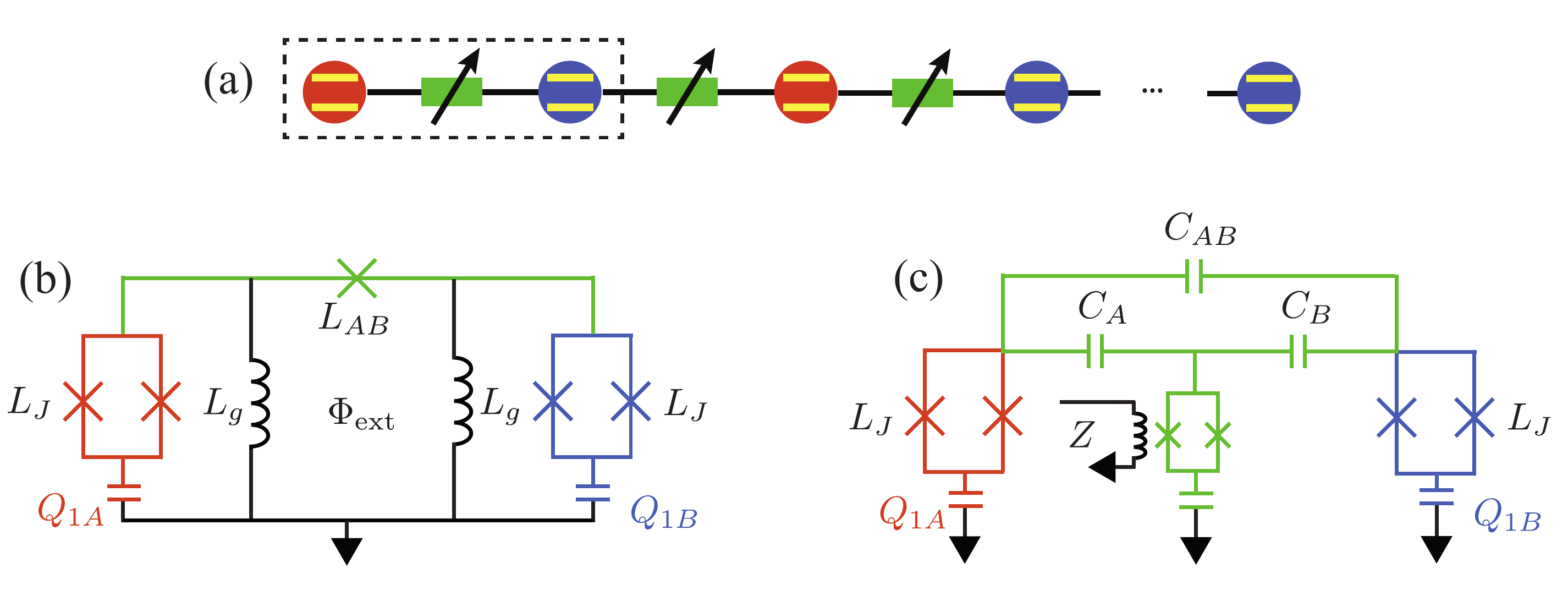}
\caption{(a) Tunable SSH qubit array. (b) Two Xmon qubits are coupled via a Josephson junction. (c) Two Xmon qubits are coupled via direct (the inductor $C_{AB}$) and indirect (the middle qubit) couplers.}\label{figureS1}
\end{figure}

\begin{figure}[!htbp]
\includegraphics[width=14cm]{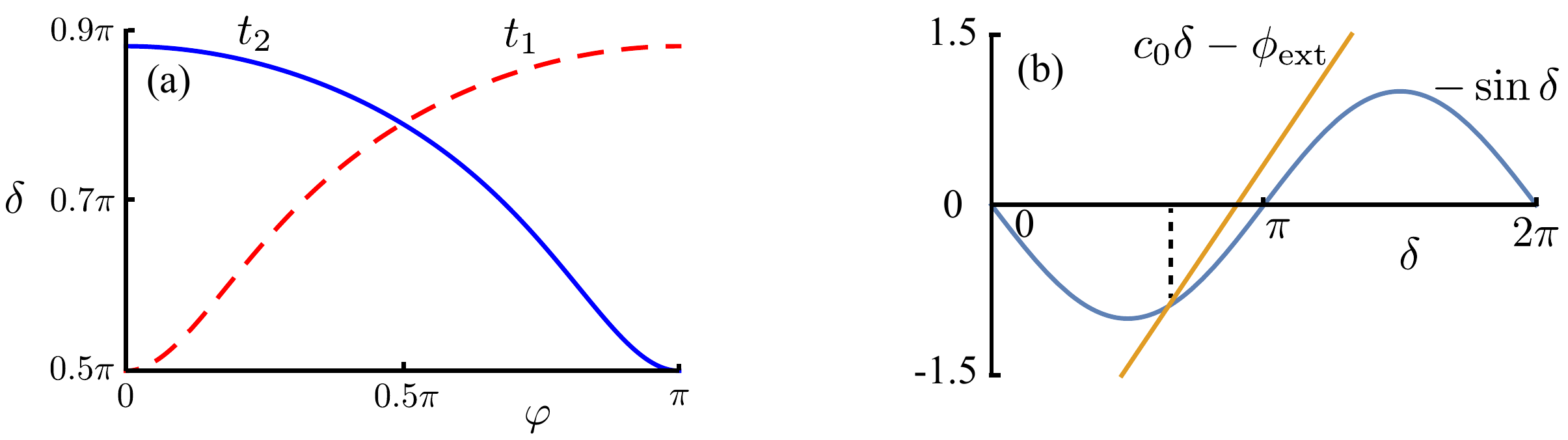}
\caption{(a) The change of $\delta$ versus $\varphi$. The red-dashed and blue-solid lines correspond to $t_1$ and $t_2$, respectively. The experimentally realizable parameters of the circuit are~\cite{Zhong2019Violating}: $L_g=0.25$ nH, $L_0=0.566$ nH and $L_J=8.34$ nH. (b) Control of $\delta$ by varying external magnetic flux.}\label{figphase}
\end{figure}

The tunable qubit-qubit couplings in superconducting quantum circuits are essential in quantum computation. It is important to realize tunable topological systems where topological phase transitions and manipulations of edge states can be achieved. Here, we propose two schemes for tunable SSH qubit arrays. In Fig.~\ref{figureS1}(a), the schematic for a 1D qubit array with tunable interactions is presented. The coupling circuits for two qubits in the first unit cell are shown in Figs.~\ref{figureS1}(b) and ~\ref{figureS1}(c). In Fig.~\ref{figureS1}(b), we consider a Josephson junction that couples two Xmon qubits. This coupling scheme has been realized in experiments for two qubits~\cite{PhysRevLett.113.220502} and a qubit array~\cite{roushan2017}. The coupling for these two qubits is~\cite{PhysRevLett.113.220502,PhysRevA.92.012320}
\begin{equation}
t=-\frac{M}{2}\frac{\omega_0}{L_J+L_g}, \label{eq2qubita}
\end{equation}
where the mutual inductance $M=L_g^2/(2L_g+L_{AB})$. The Josephson inductance is $L_{AB}=\Phi_0/(2\pi I_0\cos\delta)=L_0/\cos\delta$ where $\Phi_0=h/2e$ is the magnetic flux quantum. Here, $I_0$ is the critical current of the coupler junction, and $\delta$ is the phase difference across the coupler junction. Therefore, the qubit-qubit coupling
\begin{equation}
t=-\frac{\omega_0}{2}\frac{L_g^2}{(L_J+L_g)(2L_g+\frac{L_0}{\cos\delta})}. \label{eq2qubitb}
\end{equation}
In the experiment~\cite{PhysRevLett.113.220502}, qubit-qubit interaction is varied from $0$ to $55$ MHz. By increasing the coupler junction critical current, it is feasible to realize larger interaction, e.g., $200$ MHz, as we considered in the main text. As shown in Ref.~\cite{PhysRevA.92.012320}, the phase difference $\delta$ across the junction can be tuned via
\begin{equation}
\frac{\Phi_0}{2\pi}\delta=\Phi_{\mathrm{ext}}-2L_g I_0 \sin\delta, \label{eqdelta}
\end{equation}
where $\Phi_{\mathrm{ext}}$ is an external magnetic flux bias, as shown in Fig.~\ref{figureS1}(b). The coupling Eq.~(\ref{eq2qubitb}) can be tuned continuously from negative to positive. In the main text, we consider qubit-qubit couplings $t_{1,2}=t_0(1\mp\cos\varphi)$. According to Eq.~(\ref{eq2qubitb}), we have
\begin{equation}
\delta=\arccos\left[-\left(\frac{2L_g}{L_0}+\frac{\omega_0 L_g^2}{2t_0L_0(L_J+L_g)}\frac{1}{1\mp\cos\varphi}\right)^{-1}\right],
\end{equation}
for $t_1$ and $t_2$, respectively. In Fig.~\ref{figphase}(a), we show the values of $\delta$ for the couplers that induce qubit-qubit interactions $t_{1,2}=t_0(1\mp\cos\varphi)$. Here, the phase difference $\delta$ should change between $0.5\pi$ and $0.9\pi$. To show how to tune $\delta$ in this range, we rewrite the Eq.~(\ref{eqdelta}) as
\begin{equation}
c_0 \delta-\phi_{\mathrm{ext}}=-\sin\delta, \label{eqdelta2}
\end{equation}
with $c_0=L_0/(2L_g)$ and $\phi_{\mathrm{ext}}=\Phi_{\mathrm{ext}}/(2L_g I_0)$. In Fig.~\ref{figphase}(b), we present the solution of Eq.~(\ref{eqdelta2}). Because $c_0>1$ for the circuit parameters we consider here, $\delta$ can be tuned between $0.5\pi$ and $0.9\pi$ by changing the external magnetic flux.

In Fig.~\ref{figureS1}(c), we consider a scheme where the coupling between two qubits is mediated by a tunable qubit~\cite{PhysRevB.73.094506,Niskanen2007,PhysRevApplied.10.054062}. When the coupler qubit is largely detuned from $1A$ and $1B$ qubits, the effective Hamiltonian is
\begin{eqnarray}
H_{2q} &=& \left(\omega_0+\frac{t_A^2}{\Delta_0}\right) \sigma_{1A}^+ \sigma_{1A}^- + \left(\omega_0 + \frac{t_B^2}{\Delta_0} \right) \sigma_{1B}^+ \sigma_{1B}^- + \left(t_{AB} + \frac{t_A t_B}{\Delta_0}\right) (\sigma_{1A}^+ \sigma_{1B}^- + \mathrm{H.c.}), \label{Htq}
\end{eqnarray}
where $t_A$ ($t_B$) is the coupling between coupler qubit and $1A$ ($1B$) qubit. The direct coupling between $1A$ and $1B$ qubits is $t_{AB}$. The frequency of the coupler qubit can be tuned. Therefore, the detuning $\Delta_0$ is tunable and can be positive or negative. We consider $t_A=t_B$. The coupling becomes
\begin{eqnarray}
t_{\mathrm{eff}}&=& t_{AB}\left(1+\frac{t_A^2}{\Delta_0 t_{AB}}\right) \nonumber \\
&=&t_0 (1-\cos\varphi).
\end{eqnarray}
where we assume $t_{AB}=t_0$ and $\cos\varphi=-\frac{t_A^2}{\Delta_0 t_{AB}}$. If we change the detuning $\Delta_0 \rightarrow -\Delta_0$, the coupling becomes $t_0(1+\cos\varphi)$.

\subsection{Couplings between qubits and transmission line resonator}
In our scheme to spectroscopically characterize the topological phase transition, we consider cavity-qubit couplings with positive and negative signs. Here, we show how to implement the coupling with different signs.

\begin{figure}[t]
\includegraphics[width=14cm]{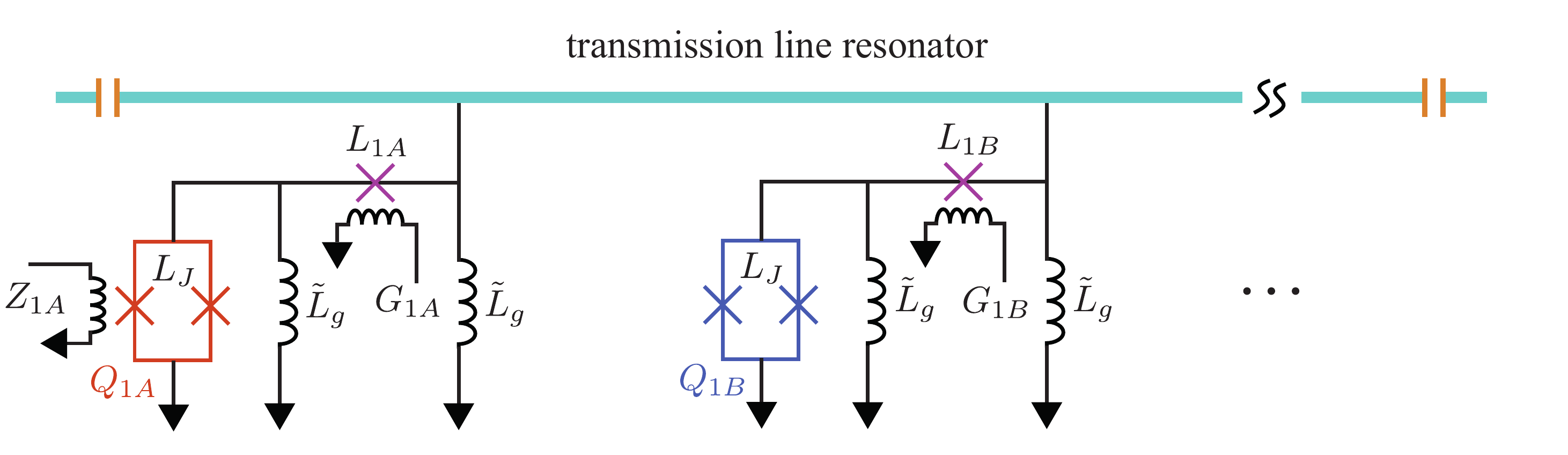}
\caption{Schematic of the circuit with tunable coupling between transmission line resonator and Xmon qubits.}\label{figureSa}
\end{figure}

In Fig.~\ref{figureSa}, we present a circuit QED scheme where qubits are coupled to a transmission line resonator. For simplicity, we consider qubit $A_1$ first. The two inductors introducing two nodes at left and right sides of the Josephson junction. This junction provides a tunable inductance $L_{1A}$ that controls the flow of current. Therefore, the coupling between qubit and transmission line resonator can be controlled. The effective mutual inductance between the qubit and transmission line resonator through the coupler is~\cite{Zhong2019Violating}
\begin{equation}
\tilde{M}=\frac{\tilde{L}_g^2}{2\tilde{L}_g+\frac{L_{1A}}{\cos\tilde{\delta}}}, \label{eqmi}
\end{equation}
where $\tilde{\delta}$ is the phase difference of the Josephson junction. The inductance of the junction is $L_{1A}=\Phi_0/(2\pi \tilde{I}_0 \cos\tilde{\delta})=\tilde{L}_0/\cos\tilde{\delta}$, where $\tilde{I}_0$ is the critical current of the junction. And $\tilde{\delta}$ can be tuned by applying a dc flux via $G_{1A}$. The interaction between qubit and transmission line resonator is
\begin{equation}
H_{\mathrm{int}}=\hbar g(\sigma_{1A}^+ \hat{a} + \hat{a}^{\dagger} \sigma_{1A}^-),
\end{equation}
with
\begin{equation}
g=-\frac{\tilde{M}}{2}\sqrt{\frac{\omega_0 \omega_c}{(\tilde{L}_g+L_{1A})(\tilde{L}_g+L_c)}},\label{eqcoupling}
\end{equation}
by considering harmonic limit and weak coupling. Here, $L_c$ is the inductance of the transmission line resonator. It can be seen from Eq.~(\ref{eqcoupling}) that the qubit-resonator coupling depends on many parameters, e.g., the inductances of circuit elements. From Eq.~(\ref{eqmi}), we can see that the sign of $\tilde{M}$ can be changed by tuning $\tilde{\delta}$. Therefore, the cavity-qubit couplings with positive/negative signs can be realized. However, as shown in Eq.~(\ref{eqcoupling}), qubit-resonator coupling is also related to frequencies of qubit and resonator. The interaction-tunable qubit array may lead to shifted frequencies of qubits, which change the qubit-resonator couplings. In the following, we analyze the effect of shifted frequency for two schemes shown in Figs.~\ref{figureS1}(b) and ~\ref{figureS1}(c), respectively.

For the coupling scheme shown in Fig.~\ref{figureS1}(b), the effective qubit inductance is
\begin{equation}
L_{q}=L_J+L_g-M.
\end{equation}
The qubit frequency $\omega_0\approx\frac{1}{\sqrt{(L_g+L_J)C_q}}$. From Eq.~(\ref{eq2qubita}), the mutual inductance is
\begin{equation}
M=-2t(L_g+L_J)\sqrt{(L_g+L_J)C_q}.
\end{equation}
The frequency of qubit is
\begin{eqnarray}
\omega_0'&=&\frac{1}{\sqrt{L_q C_q}} \nonumber \\
&\approx& \omega_0(1-t\sqrt{(L_g+L_J)C_q}).
\end{eqnarray}
Therefore, the coupler yields frequency shift
\begin{equation}
\delta\omega= -t.
\end{equation}
Therefore, the frequency shift of qubit depends on the tunable parameter of the coupler. In Ref.~\cite{PhysRevLett.113.220502}, this frequency shift is compensated by applying a $Z$ control of the qubit. In the SSH qubit array, the frequency shift for qubits that have $t_1$ and $t_2$ couplings to their neighboring qubits is $-2t_0$. This means that the frequency shifts for these qubits are independent of the tunable parameter. From Eq.~(\ref{eqcoupling}), the change of qubit-cavity coupling is
\begin{equation}
\delta g=-\frac{t_0 g_0}{\omega_0}.
\end{equation}
For the parameters considered in the main text, $\delta g=-0.083\times 2\pi$ MHz, which is small comparing with $g_0=5\times 2\pi$ MHz.

For the coupling scheme shown in Fig.~\ref{figureS1}(c), an auxiliary qubit is used as a coupler and induces frequency shifts to two qubits (see Eq.~(\ref{Htq})). However, for a qubit that has $t_{1,2}=t_0(1\mp\cos\varphi)$ to its neighboring qubits, its effective frequency shift is zero. Therefore, in the SSH qubit arrays with tunable qubit-qubit interactions shown in Figs.~\ref{figureS1}(b) and ~\ref{figureS1}(c), the frequencies of qubits are not changed as we tune $\varphi$. Hence, the qubit-resonator coupling (see Eq.~\ref{eqcoupling}) is robust to the tunable SSH qubit array. However, due to open boundary conditions, two qubits at ends of the array have different frequency shifts comparing to other qubits. We can use the $Z$ bias in Fig.~\ref{figureSa} to compensate these frequency shifts such that all the qubits in the array have the same frequency.

\subsection{Couplings between eigenmodes and transmission line resonator}
The couplings between cavity and qubits are described by
\begin{equation}
H_{I}=\sum_{i=1,\mu=A,B}^{i=N} g_{i\mu}\sigma_{i\mu}^+\hat{a} + g_{i\mu}^{*}\hat{a}^{\dagger}\sigma_{i\mu}^-.
\end{equation}
We consider homogeneous coupling, i.e., $g_{i\mu}=g_0$. We consider the single-excitation subspace $|\mathcal{A}_i\rangle=\sigma_{iA}^+|G\rangle$ and $|\mathcal{B}_i\rangle=\sigma_{iB}^+|G\rangle$ where $|G\rangle$ is the ground state of the qubit array. Therefore, we have $\langle G|\sigma_{iA}^-|\mathcal{A}_i\rangle=1$, $\langle \mathcal{A}_i|\sigma_{iA}^+|G\rangle=1$, $\langle G|\sigma_{iB}^-|\mathcal{B}_i\rangle=1$ and $\langle \mathcal{B}_i|\sigma_{iB}^+|G\rangle=1$. In the single-excitation subspace, the operators of qubits can be written as superposition of eigenmodes, $\sigma_{iA}^{+}\rightarrow|\mathcal{A}_i\rangle\langle G|=\sum_j \xi_{2i-1,j} \Psi_j^{+}$ and $\sigma_{iB}^{+}\rightarrow|\mathcal{B}_i\rangle\langle G|=\sum_j \xi_{2i,j} \Psi_j^{+}$. Then qubit array and cavity interaction Hamiltonian can be written as
\begin{eqnarray}
H_I&=& \sum_j g_0 \Psi_j^+\hat{a}\sum_{i=1}^{N} (\xi_{2i-1,j}+\xi_{2i,j})  + \mathrm{H.c.} \nonumber \\
&=& \sum_j \xi_j g_0 \Psi_j^+ \hat{a} + \mathrm{H.c.} \nonumber \\
&=& \sum_j \tilde{\xi}_j \Psi_j^+ \hat{a} + \mathrm{H.c.},
\end{eqnarray}
where $\tilde{\xi}_j=\xi_jg_0$. The cavity-eigenmode coupling coefficient $\xi_j$ is the summation of all the components of the $j$th eigenstate, i.e.,
\begin{equation}
\xi_j=\sum_{i=1}^{N} (\xi_{2i-1,j}+\xi_{2i,j}). \label{eqxi}
\end{equation}
The wavefunctions for left and right edge states are
\begin{eqnarray}
|\Psi_L\rangle &=& \frac{1}{\sqrt{\mathcal{N}_L}}\sum_{i=1}^N \left(-\frac{t_1}{t_2}\right)^{i-1}|\mathcal{A}_i\rangle, \label{wfledge} \\
|\Psi_R\rangle &=& \frac{1}{\sqrt{\mathcal{N}_R}}\sum_{i=1}^N \left(-\frac{t_1}{t_2}\right)^{N-i}|\mathcal{B}_i\rangle. \label{wfredge}
\end{eqnarray}
where $\mathcal{N}_L$ and $\mathcal{N}_R$ are the normalization factors ($L,R\in \{N,N+1\}$). Therefore, $\xi_{2i-1,L}= \frac{1}{\sqrt{\mathcal{N}_L}} \big(-\frac{t_1}{t_2}\big)^{i-1}$ and $\xi_{2i,L}=0$ for the left edge state, $\xi_{2i-1,R}=0$ and $\xi_{2i,R}= \frac{1}{\sqrt{\mathcal{N}_R}}\big(-\frac{t_1}{t_2}\big)^{N-i}$ for the right edge state.
\begin{figure}[b]
\includegraphics[width=16cm]{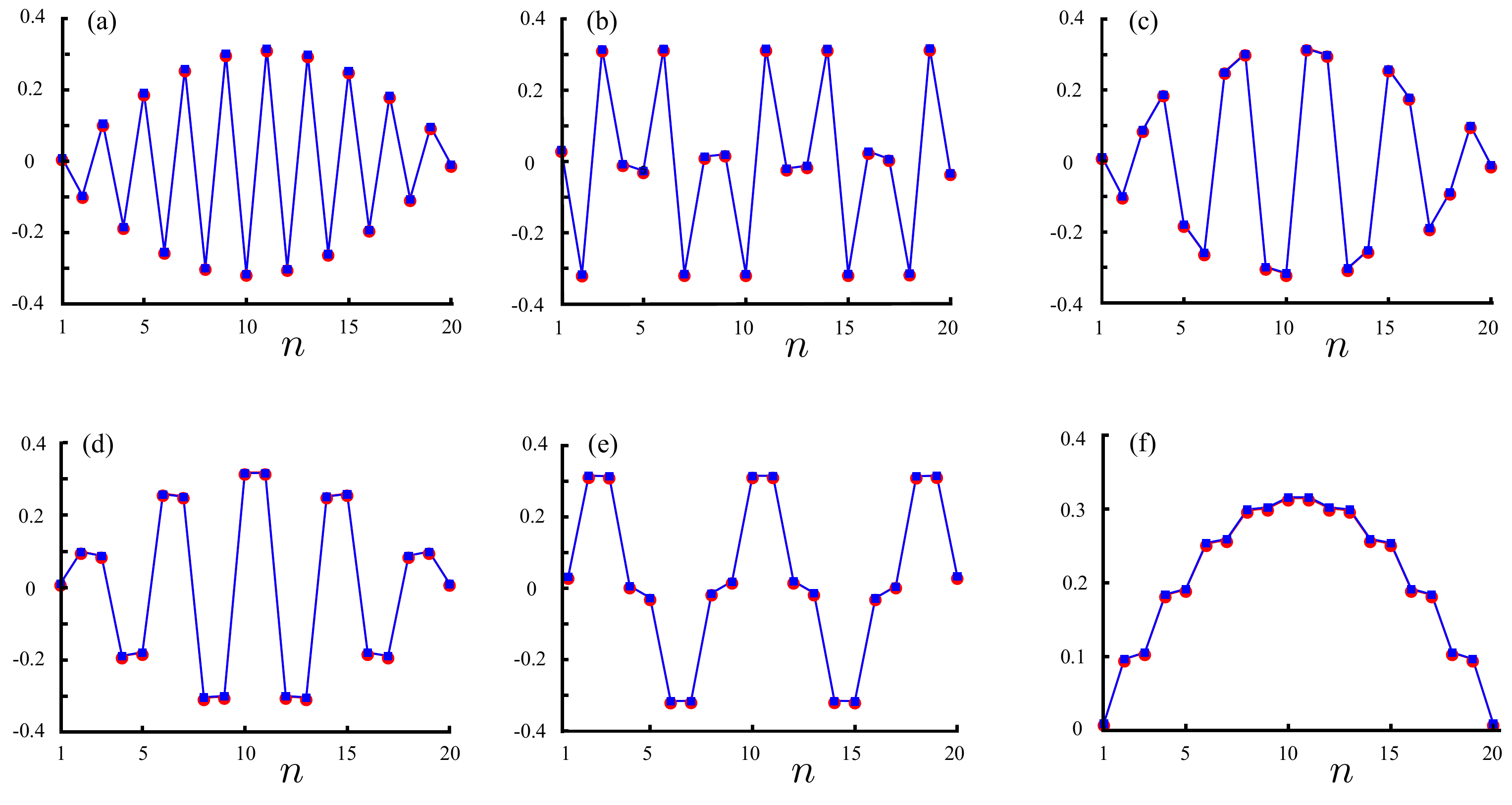}
\caption{Wavefunctions of bulk states with (a) $j=1$, (b) $j=5$, (c) $j=9$, (d) $j=12$, (e) $j=16$ and (f) $j=20$. We here consider the qubit array with $10$ unit cells. The red circles and blue squares represent numerical and analytical results of the wavefunctions.}\label{bulk}
\end{figure}

Now we consider wavefunctions of bulk states. The wavefunctions of bulk states in the lower and upper bands of bulk states have different forms~\cite{PhysRevB.84.195452}, i.e.,
\begin{equation}
|\Psi_j\rangle = \frac{1}{\sqrt{2N}} \sum_{i=1}^{N} \sin (ik_j-\phi(k_j))|\mathcal{A}_i\rangle - \sin (ik_j)|\mathcal{B}_i\rangle, \quad\quad\quad \mathrm{for} \quad j\in [1,N-1] \label{Eqbw1}
\end{equation}
and
\begin{equation}
|\Psi_j\rangle = \frac{1}{\sqrt{2N}} \sum_{i=1}^{N} \sin (ik_j-\phi(k_j))|\mathcal{A}_i\rangle + \sin (ik_j)|\mathcal{B}_i\rangle, \quad\quad\quad \mathrm{for} \quad j\in [N+2, 2N]  \label{Eqbw2}
\end{equation}
where
\begin{equation}
\phi(k)=\arccot\left(\frac{t_1}{t_2 \sin k}+\cot k\right).
\end{equation}
Note that $j=N$ and $N+1$ represent edge states. Therefore, $\xi_{2i-1,j}=\sin (ik_j-\phi(k_j))$ and $\xi_{2i,j}=\mp\sin (ik_j)$ for lower and upper bulk bands. The boundary condition $\langle \mathcal{A}_{N+1} |\Psi_j\rangle=0$ imposes the quantization condition $k(N+1)-\phi(k)=\tau\pi$
with $\tau\in[1,N-1]$ in the topological phase~\cite{PhysRevB.84.195452}. Here, we consider the lower and upper bulk bands separately.

Case (1): lower band with $j\in [1,N-1]$.
Therefore, we have
\begin{equation}
k_j(N+1)-\phi(k_j)=\tau_j\pi,
\end{equation}
with $\tau_j=j$. The above equation is solved in the range $0<k_j<\pi$. The solutions of $k_j$ span a space $\mathcal{S}$ with size $N-1$, i.e., $\mathcal{S}(j)=k_j$. In Figs.~\ref{bulk}(a)-\ref{bulk}(c), we show the analytical and numerical wavefunctions of some bulk states in the lower band. From Eq.~(\ref{Eqbw1}), we can know parity properties of bulk states, i.e., the wavefunction relation for pairs of qubits $A_i\leftrightarrow B_{N+1-i}$ and $B_i\leftrightarrow A_{N+1-i}$. The component in qubit $B_{N+1-i}$ is
\begin{eqnarray}
\xi_{2N+2-2i,j}&=&\frac{-1}{\sqrt{2N}}\sin [(N+1-i)k_j] \nonumber \\
&=&\frac{-1}{\sqrt{2N}}\sin [\tau_j\pi-(ik_j-\phi(k_j))] \nonumber  \\
&=&\frac{1}{\sqrt{2N}}(-1)^{\tau_j}\sin (ik_j-\phi(k_j)) \nonumber \\
&=&(-1)^{j}\xi_{2i-1,j}. \label{Eqparity1}
\end{eqnarray}

Case (2): upper band with $j\in [N+2,2N]$.
\begin{equation}
k_j(N+1)-\phi(k_j)=\tau'_j\pi,
\end{equation}
with $\tau'_j=\tau_{2N+1-j}$ and $k_j=\mathcal{S}(2N+1-j)$. In Figs.~\ref{bulk}(d)-\ref{bulk}(f), we show the analytical and numerical wavefunctions of some bulk states in the upper band. From Eq.~(\ref{Eqbw2}), we can know the parity properties of bulk states. The component in qubit $B_{N+1-i}$ is
\begin{eqnarray}
\xi_{2N+2-2i,j}&=&\frac{1}{\sqrt{2N}}\sin [(N+1-i)k_j] \nonumber \\
&=&\frac{1}{\sqrt{2N}}\sin [\tau'_j\pi-(ik_j-\phi(k_j))] \nonumber  \\
&=&\frac{1}{\sqrt{2N}}(-1)^{\tau'_j+1}\sin (ik_j-\phi(k_j)) \nonumber \\
&=&(-1)^{j} \xi_{2i-1,j}. \label{Eqparity2}
\end{eqnarray}

From Eqs.~(\ref{Eqbw1}),~(\ref{Eqbw2}),~(\ref{Eqparity1}) and (\ref{Eqparity2}), we know that in the $j$th bulk state, the qubits $A_i$ and $B_{N+1-i}$ have the same (opposite) component when $j$ is even (odd). In other words, the $j$th bulk states with $j$ being odd (even) have the odd (even) parity, for $j\in [1,N-1]\cup [N+2,2N]$. Because $\xi_j$ is the summation of all the components of the $j$th eigenstate, $\xi_j$ is zero for bulk states with odd parity. However, for bulk states with even parity, $\xi_j$ can be expressed
\begin{equation}
\xi_j=\sqrt{\frac{2}{N}}\sum_{i=1}^N\sin (ik_j-\phi(k_j)).
\end{equation}
Hence, the rescaling factor is
\begin{eqnarray}
\epsilon_j&=&\frac{\xi_j}{\sqrt{2N}} \nonumber \\
&=&\frac{1}{N}\sum_{i=1}^N\sin (ik_j-\phi(k_j)).
\end{eqnarray}
Now, we consider the couplings between edge modes and cavity. By considering $t_1/t_2=(1-\cos\varphi)/(1+\cos\varphi)=\tan^2 \frac{\varphi}{2}$, from Eq.~(\ref{wfledge}) we can have
\begin{eqnarray}
\xi_L&=&\frac{1}{\sqrt{\mathcal{N}_L}}\left(1-\tan^2\frac{\varphi}{2}+\tan^4\frac{\varphi}{2}-\tan^6\frac{\varphi}{2}+\ldots\right) \nonumber \\
&=&\frac{1}{\sqrt{\mathcal{N}_L}} \frac{1}{1+\tan^2\frac{\varphi}{2}}.
\end{eqnarray}
where
\begin{eqnarray}
\mathcal{N}_L&=&1+|\frac{t_1}{t_2}|^2+|\frac{t_1}{t_2}|^3+\ldots \nonumber \\
&=&\frac{1}{1-\tan^4\frac{\varphi}{2}}.
\end{eqnarray}
Therefore, we obtain
\begin{equation}
\xi_L=\sqrt{\frac{1-\tan^2\frac{\varphi}{2}}{1+\tan^2\frac{\varphi}{2}}}=\sqrt{\cos\varphi}. \label{eqcl}
\end{equation}
Similarly, we can obtain $\xi_R=\sqrt{\cos\varphi}$. As $\varphi$ increases, the separated edge states become hybridized. As long as the hybridization is not strong, i.e., the splitting between hybridized edge states is small, we can write the hybridized edge states as $|\Psi_{\pm}\rangle=\frac{1}{\sqrt{2}}(|\Psi_L\rangle \pm |\Psi_R\rangle)$. Therefore, we have
\begin{equation}
\xi_{2i-1,j}=\frac{1}{\sqrt{2\mathcal{N}_L}}(-\frac{t_1}{t_2})^{i-1}, \quad \xi_{2i,j}=\frac{1}{\sqrt{2\mathcal{N}_L}}(-\frac{t_1}{t_2})^{N-i},
\end{equation}
for $|\Psi_+\rangle$, and
\begin{equation}
\xi_{2i-1,j}=\frac{1}{\sqrt{2\mathcal{N}_L}}(-\frac{t_1}{t_2})^{i-1}, \quad \xi_{2i,j}=-\frac{1}{\sqrt{2\mathcal{N}_L}}(-\frac{t_1}{t_2})^{N-i},
\end{equation}
for $|\Psi_-\rangle$. For the state $|\Psi_+\rangle$, if we exchange the position of the left and the right edge states, then we have $|\Psi_{+,\mathrm{exch}}\rangle=\frac{1}{\sqrt{2}}(|\Psi_L\rangle + |\Psi_R\rangle)=|\Psi_+\rangle$. In this sense, we say that $|\Psi_+\rangle$ has even parity. However, the state $|\Psi_-\rangle$ will change sign if the left and right edge states exchange their position, that is, $|\Psi_{-,\mathrm{exch}}\rangle=-\frac{1}{\sqrt{2}}(|\Psi_L\rangle - |\Psi_R\rangle)=-|\Psi_-\rangle$. Then we say that the state $|\Psi_-\rangle$ has odd parity. If the state $|\Psi_+\rangle$ is coupled to the cavity, according to Eq.~(\ref{eqxi}), the coupling coefficient is
\begin{eqnarray}
\xi_+&=&\frac{1}{\sqrt{2\mathcal{N}_L}}2\left(1-\tan^2\frac{\varphi}{2}+\tan^4\frac{\varphi}{2}-\tan^6\frac{\varphi}{2}+\ldots\right)\nonumber\\
&=&\sqrt{\frac{2}{\mathcal{N}_L}}\frac{1}{1+\tan^2\frac{\varphi}{2}}\nonumber \\
&=&\sqrt{2\cos\varphi}, \label{eqsymmetric}
\end{eqnarray}
which is nonzero in the topological phase. That is, the state with even parity is coupled to the cavity. However, if the state $|\Psi_-\rangle$ is coupled to the cavity, the summation of all the components of this state is zero. Therefore, $\xi_{-}=0$. That is, the state $|\Psi_-\rangle$ is decoupled from the cavity.

\section{Cavity spectroscopy of topological qubit arrays}
The Hamiltonian of the qubit array with SSH interactions is
\begin{equation}
H_{\mathrm{SSH}}= \sum_{i=1}^{N} \omega_0\big( \sigma_{iA}^+ \sigma_{iA}^- + \sigma_{iB}^+ \sigma_{iB}^-\big) + \sum_{i=1}^{N} \big(  t_{1} \sigma_{iA}^+ \sigma_{iB}^- + t_2 \sigma_{i+1A}^+ \sigma_{iB}^- + \mathrm{H.c.} \big).
\end{equation}
We denote $|\mathcal{A}_i\rangle= \sigma_{iA}^+ |G\rangle$ and $|\mathcal{B}_i\rangle= \sigma_{iB}^+ |G\rangle$ with $|G\rangle$ being the ground state of the qubit array. Then we have
\begin{eqnarray}
\langle \mathcal{A}_i|\sigma_{iA}^+ \sigma_{iB}^-|\mathcal{B}_i\rangle&=&\langle G|G\rangle \nonumber \\
&=&1.
\end{eqnarray}
and similarly $\langle \mathcal{A}_{i+1}|\sigma_{i+1A}^+ \sigma_{iB}^-|\mathcal{B}_i\rangle=1$. Therefore, the system in the single-excitation subspace $\{|\mathcal{A}_i\rangle, |\mathcal{B}_i\rangle\}$ is described by
\begin{equation}
\mathcal{H}_{\mathrm{SSH}}=\sum_{i=1}^{N} \omega_0\big( |\mathcal{A}_i\rangle\langle\mathcal{A}_i| + |\mathcal{B}_{i}\rangle\langle\mathcal{B}_i|\big) + \sum_{i=1}^{N} \big(t_{1} |\mathcal{A}_i\rangle\langle\mathcal{B}_i| + t_2 |\mathcal{A}_{i+1}\rangle\langle\mathcal{B}_i| + \mathrm{H.c.} \big).
\end{equation}
\begin{figure}[t]
\includegraphics[width=14cm]{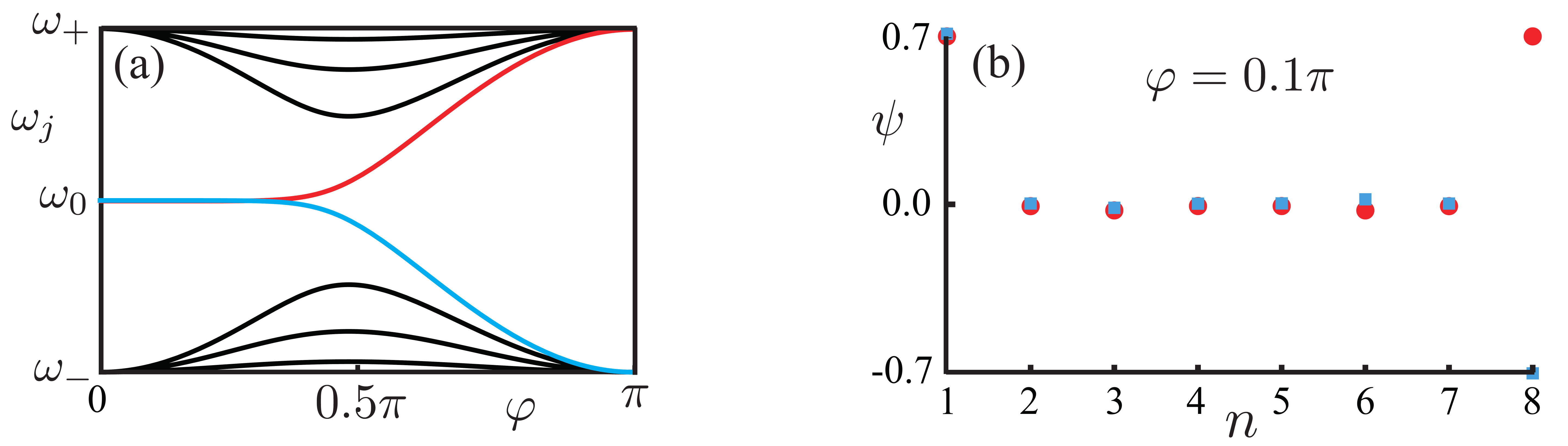}
\caption{(a) Energy spectrum for the array with $8$ qubits. Black-solid lines represent the bulk states. Red and blue curves show the edge states and their transitions to bulk states across the critical point at $\varphi=0.5\pi$. As the same to Fig.~1(d), $\omega_0$ represents the frequency of qubits, and $\omega_{\pm}=\omega_0\pm 2 t_0$. (b) Wavefunctions of edge states at $\varphi=0.1\pi$. Red dots and blue squares denote hybridized edge states with even and odd parities, respectively. }\label{spectrum}
\end{figure}
In Fig.~\ref{spectrum}(a), we show the energy spectrum of SSH array with $8$ qubits. The red and blue curves represent edge states ($\varphi<\frac{\pi}{2}$) and transition to bulk states ($\varphi>\frac{\pi}{2}$). And the wavefunctions corresponding to two edge states at $\varphi=0.1\pi$ are shown in Fig.~\ref{spectrum}(b). Due to finite size of the system, edge states are hybridized with even and odd parities. In solid state systems, the edge states, e.g., Majorana fermions, and topological phase transitions are probed via electronic transport (see Refs. [13,49-51] in the main text). In the topological qubit array as we considered here, cavity spectroscopy can be used to observe the topological phase transition.

We consider that the SSH qubit array is coupled to a cavity, as shown in Fig.~1(b). The master equation describing the whole system is
\begin{equation}
\dot{\rho} = -\frac{i}{\hbar}\Big[H+i \hbar\eta(\hat{a}^{\dagger}e^{-i\omega_l t} - \hat{a}e^{i \omega_l t}), \rho\Big] + \mathcal{L}_a[\rho] + \mathcal{L}_c[\rho],
\end{equation}
where $H$ is the Hamiltonian containing the cavity, qubit array and their coupling, as described by Eq.~(1) in the main text. And $\eta$ and $\omega_l$ are the driving strength and frequency of the cavity. The dissipation terms for the qubit array and cavity are respective
\begin{equation}
\mathcal{L}_a[\rho] = \sum_{i,\mu=A,B} \gamma_{i\mu} (\sigma_{i\mu}^- \rho \sigma_{i\mu}^+ - \frac{1}{2}\sigma_{i\mu}^+ \sigma_{i\mu}^- \rho - \frac{1}{2}\rho \sigma_{i\mu}^+ \sigma_{i\mu}^-),
\end{equation}
and
\begin{equation}
\mathcal{L}_c[\rho] = \kappa(\hat{a} \rho \hat{a}^{\dagger}-\frac{1}{2}\hat{a}^{\dagger} \hat{a} \rho -\frac{1}{2}\rho \hat{a}^{\dagger} \hat{a}).
\end{equation}
\begin{figure}[t]
\includegraphics[width=16cm]{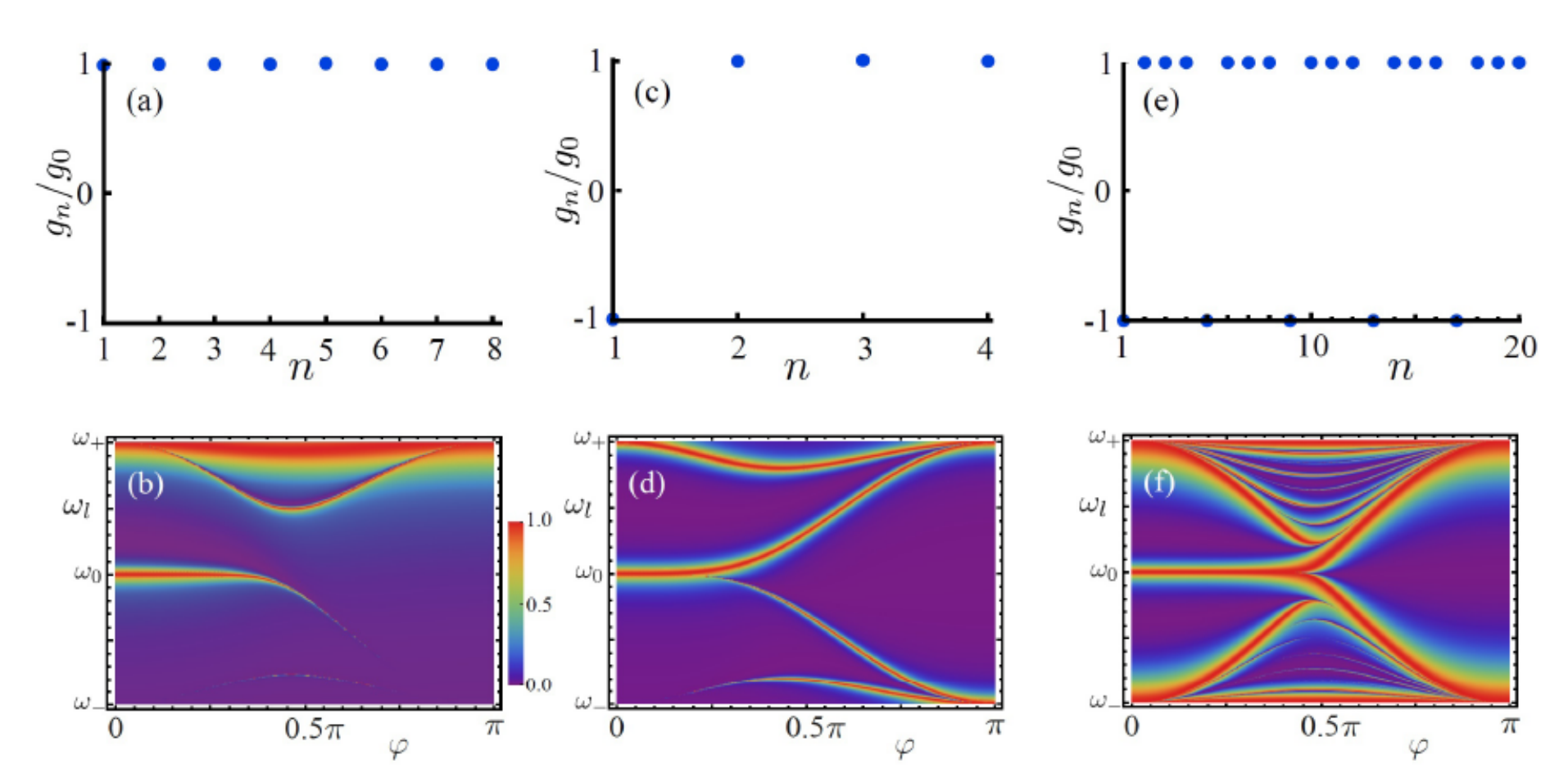}
\caption{Cavity-qubit couplings and the corresponding cavity spectroscopy of the qubit array. (a) All the qubits have the same coupling to the cavity. (b) cavity spectroscopy with the couplings (a). The size of the qubit array is $N=4$ unit cells. (c) Inhomogeneous couplings between cavity and qubits for the array with $2$ unit cells and (d) cavity spectroscopy. (e) Inhomogeneous couplings between cavity and qubits for the array with $N=10$ unit cells and (f) cavity spectroscopy.}\label{spectroscopy}
\end{figure}
We consider low-excitation limit, i.e., $\langle\sigma_{i\alpha}^+\sigma_{i\alpha}^-\rangle\approx 0$ (with $\alpha=A,B$), which can be realized by a weak external probe field~\cite{PhysRevLett.68.1132,PhysRevA.75.053823,Tiecke2014}. From the master equation, we obtain the equations
\begin{eqnarray}
\big\langle \frac{d}{dt} \hat{a}\big\rangle &=& -(\frac{\kappa}{2} + i \Delta_c) \langle \hat{a} \rangle -i \bm{g}^\intercal \langle \bm{\sigma} \rangle + \eta,  \\
\big\langle \frac{d}{dt}\bm{\sigma} \big\rangle &=& -i (\Delta_q + \bm{D} -i \frac{\bm{\Gamma}}{2}) \langle \bm{\sigma} \rangle -i \bm{g} \langle \hat{a} \rangle,
\end{eqnarray}
with $\Delta_c = \omega_c - \omega_l$, $\Delta_q = \omega_0 - \omega_l$, $\bm{g}=(g_{1A},g_{1B},g_{2A},g_{2B}, \cdots)$, $\langle \bm{\sigma} \rangle=(\langle \sigma_{1A}^- \rangle, \langle \sigma_{1B}^- \rangle, \langle \sigma_{2A}^- \rangle, \langle \sigma_{2B}^- \rangle, \cdots)^\intercal$, $\bm{\Gamma}=\mathrm{diag}(\gamma_{1A},\gamma_{1B},\gamma_{2A},\gamma_{2B},\cdots)$ and
\begin{eqnarray}
\bm{D} = \left(
                \begin{array}{ccccc}
                  0 & t_1 & 0 & 0 & 0 \\
                  t_1 & 0 & t_2 & 0 & 0 \\
                  0 & t_2 & 0 & t_1 & 0 \\
                  0 & 0 & t_1 & 0 & t_2 \\
                  0 & 0 & 0 & t_2 & \ddots \\
                \end{array}
              \right).
\end{eqnarray}
In steady state, the photon reflection
\begin{equation}
R=1-\Big|\frac{\frac{\kappa}{2}}{\frac{\kappa}{2} + i \Delta_c -i \sum_{nm} g_n g_m[(\Delta_q + \bm{D} -i \frac{\bm{\Gamma}}{2})^{-1}]_{nm}}\Big|^2.
\end{equation}
In Fig.~1(d) of the main text, we show the reflection spectrum of the qubit array with cavity-qubit couplings $\bm{g}=g_0(-1,1,1,1,-1,1,1,1)$. These couplings lead to spectroscopic signature of topological phase transition. In order to
draw a comparison, we consider homogeneous couplings between cavity and qubits in the array with the same size, i.e., $N=4$ unit cells, as shown in Fig.~\ref{spectroscopy}(a). The corresponding reflection spectrum is demonstrated in Fig.~\ref{spectroscopy}(b). The bulk states with even parity in the higher energy band can be observed. However, the signature for the edge-bulk transition is inhibited. This is because of the parity properties of the edge and bulk states. The couplings in Figs.~\ref{spectroscopy}(c) and \ref{spectroscopy}(e) produce the spectroscopic measurements, shown in Figs.~\ref{spectroscopy}(d) and \ref{spectroscopy}(f), for the qubit arrays with $2$ and $10$ unit cells, respectively. The edge states and their transitions to bulk states can be observed. Therefore, the cavity-qubit couplings are important to uncover topological phase transition in the cavity spectroscopy approach.

\begin{figure}[b]
\includegraphics[width=14cm]{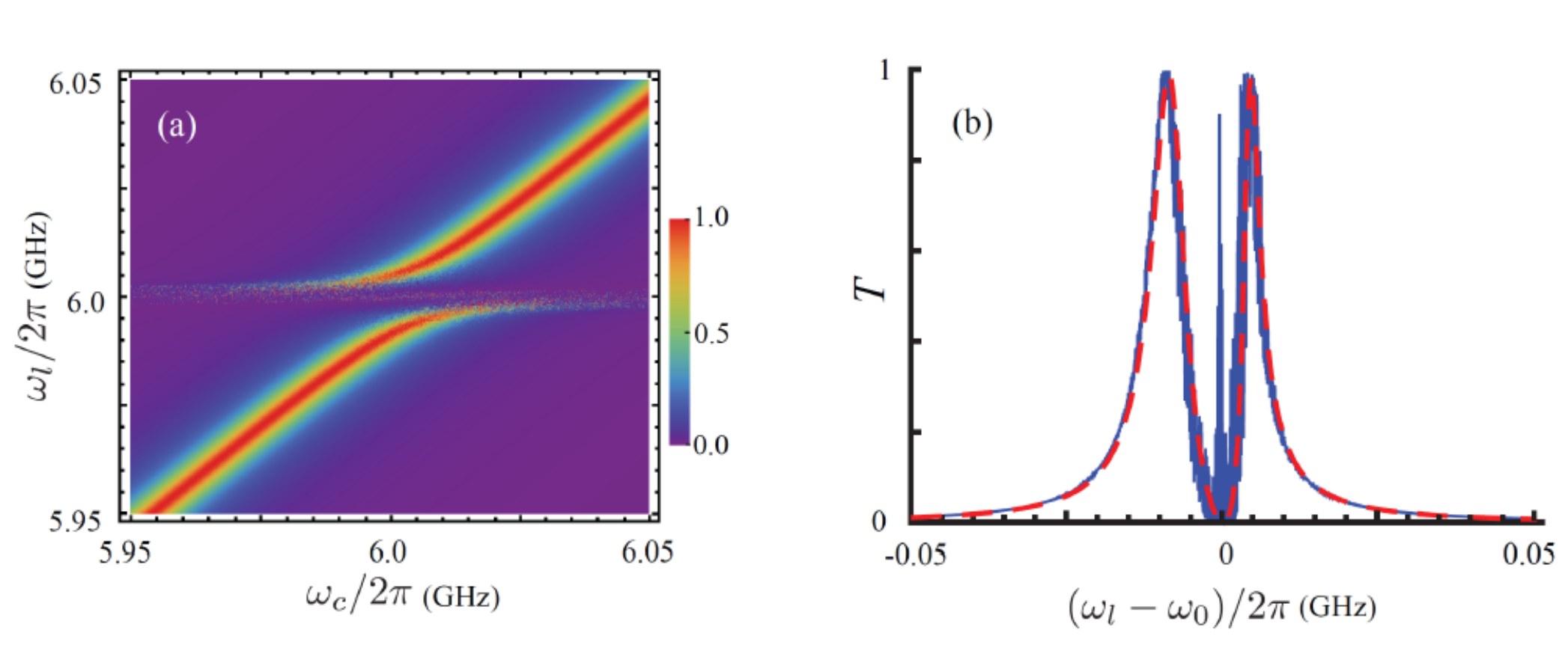}
\caption{(a) The effect of disorder in Rabi splitting of edge states. (b) Transmission spectrum with $\omega_c=\omega_0$. The red-dashed and blue-solid lines correspond to disorder strength $\epsilon=0$ and $\epsilon=2\times2\pi$ MHz, respectively. Here we consider $N=18, \varphi=0.2\pi$. Other parameters are as the same as Fig.~1(d) in the main text.}\label{figdisorder}
\end{figure}

The disorder of qubits' frequencies is studied in the spectroscopic measurement, as shown in Fig.~\ref{figdisorder}(a). The frequencies of qubits are $\omega_{0}+\epsilon_{i\alpha}$ ($\alpha=A,B$), where $\epsilon_{i\alpha}$ are randomly distributed $\epsilon_{i\alpha}\in[-\epsilon,\epsilon]$. Here, $\epsilon$ represents the strength of the disorder. In Fig.~\ref{figdisorder}(b), we show the transmission with cavity-qubit resonance, i.e., $\omega_c=\omega_0$. The red-dashed line represents the transmission spectrum with no disorder. The distance between two peaks is the Rabi splitting produced by edge states. When the disorder is considered, the Rabi splitting of edge states can still be resolved, shown by the blue-solid line. Interestingly, the disorder induces transparency at $\omega_l=\omega_c=\omega_0$.

In the cavity QED with a single qubit, the condition to resolve the Rabi splitting is $2g_0>(\kappa+\gamma)/2$ where $g_0$ is the cavity-qubit coupling; $\kappa$ and $\gamma$ are respective the decays of cavity and qubit. In the Rabi splitting for edge states, the effective coupling between cavity and edge states is $\sqrt{2\cos\varphi}g_0$, which is about $1.27g_0$ for $\varphi=0.2\pi$. Therefore, the condition to resolve the Rabi splitting of edge states is $2.54g_0>(\kappa+\gamma)/2$ for $\varphi=0.2\pi$. Here, the decay for unhybridized edge states is $\gamma$. For the parameters we considered in the main text, the condition of resolved Rabi splitting is satisfied, and thus the Rabi splitting of edge states can be resolved (see Figs.~\ref{figdisorder}(a) and \ref{figdisorder}(b)). However, for the reason that the effective coupling between edge states and the cavity is $\sqrt{2\cos\varphi}g_0$, the Rabi splitting for edge states is easier to be resolved than the one for single qubit when $\varphi<\pi/3$.

\section{Quantum dynamics protected by topological bandgap}
When the cavity is detuned with the topological qubit array, effective couplings among all the qubits can be obtained. Now we show how to obtain the effective Hamiltonian. The Hamiltonian for the cavity and SSH qubit array is
\begin{equation}
H/\hbar = \omega_c \hat{a}^{\dagger} \hat{a} + \sum_{i=1,\mu=A,B}^{i=N} \big(\omega_0 \sigma_{i\mu}^+ \sigma_{i\mu}^- + g_{i\mu} \sigma_{i\mu}^+ \hat{a} +   g_{i\mu}^* \hat{a}^{\dagger} \sigma_{i\mu}^- \big)
+\sum_{i=1}^{i=N} \big(t_{1} \sigma_{iA}^+ \sigma_{iB}^- + t_2 \sigma_{i+1A}^+ \sigma_{iB}^- + \mathrm{H.c.} \big).
\end{equation}
We consider homogenous couplings between cavity and qubits, i.e., $g_{i\mu}\equiv g_0$. And in a rotating frame with $H_{\mathrm{rot}}=\hbar\omega_c(\hat{a}^{\dagger} \hat{a}+\sum_{i,\mu}\sigma_{i\mu}^+ \sigma_{i\mu}^- )$, the Hamiltonian can be rewritten as
\begin{equation}
H'/\hbar=\Delta_0 \sum_{i,\mu}\sigma_{i\mu}^+ \sigma_{i\mu}^- +g_0\sum_{i,\mu}(\sigma_{i\mu}^+\hat{a}+\hat{a}^{\dagger}\sigma_{i\mu}^-)
+\sum_{i=1}^{i=N} \big(t_{1} \sigma_{iA}^+ \sigma_{iB}^- + t_2 \sigma_{i+1A}^+ \sigma_{iB}^- + \mathrm{H.c.} \big),
\end{equation}
with $\Delta_0=\omega_0-\omega_c$. If $g_0\ll\Delta_0$, we can make a unitary transformation with
\begin{equation}
U=\exp\Big[\frac{g_0}{\Delta_0}\sum_{i\mu}(\sigma_{i\mu}^+\hat{a}-\hat{a}^{\dagger}\sigma_{i\mu}^-)\Big],
\end{equation}
and obtain an effective Hamiltonian
\begin{equation}
H_{\mathrm{eff}}=\Delta_0 \sum_{i,\mu}\sigma_{i\mu}^+ \sigma_{i\mu}^-  + \frac{g_0^2}{\Delta_0}\sum_{i,j,\mu,\nu} \sigma_{i\mu}^+\sigma_{j\nu}^- +\sum_{i=1}^{i=N} \big(t_{1} \sigma_{iA}^+ \sigma_{iB}^- + t_2 \sigma_{i+1A}^+ \sigma_{iB}^- + \mathrm{H.c.} \big).
\end{equation}
The second term contains the Lamb shifts and exchange interactions between qubits which are produced by the cavity. In this work, when we study the cavity-mediated interactions between qubits, the condition $g_0/\Delta_0=0.1$ is considered. As we show in the main text, such global coupling has different dynamical effects in topological and non-topological regimes, depending on the bandgap. In the topological phase, cavity mediates the coupling between two edge states. In the non-topological phase, the cavity-induced couplings between bulk states are much smaller than the SSH interactions. The edge-state coupling has potential applications in quantum information processing. For example, quantum states can be transferred between two edge states, yielding topology-protected state transfer. Moreover, this nonlocal coupling can lead to interesting quantum optical phenomena, and is promising for quantum control with edge states.

\subsection{Rabi oscillation between edge states}
The effective Hamiltonian mediated by the cavity can be written as
\begin{eqnarray}
\bar{H}/\hbar=\sum_{j} \Big(\Delta_{j} + \frac{\tilde{\xi}^2_{j}}{\Delta_{j}}\Big)|\Psi_{j}\rangle \langle \Psi_{j}| +  \sum_{j\neq k} \frac{\tilde{\xi}_j \tilde{\xi}_k}{2}\Big( \frac{1}{\Delta_j} + \frac{1}{\Delta_k} \Big) (|\Psi_j\rangle \langle \Psi_k| + \mathrm{H.c.}). \label{Hcoupling}
\end{eqnarray}
The edge states without hybridization have effective coupling $J=\tilde{\xi}_{N} \tilde{\xi}_{N+1}/\Delta_0$ which comes from the second term in Eq.~(\ref{Hcoupling}). From Eq.~(\ref{eqcl}), we can obtain
\begin{equation}
J=\cos\varphi \frac{g_0^2}{\Delta_0}. \label{eqec1}
\end{equation}
However, when the edge states are hybridized, the edge state with odd parity is not coupled to the cavity, i.e., the second term in Eq.~(\ref{Hcoupling}) is vanishing. Therefore, the Hamiltonian Eq.~(\ref{Hcoupling}) becomes
\begin{eqnarray}
\bar{H}'/\hbar&=&\Big(\Delta_{0} + \frac{\tilde{\xi}^2_{+}}{\Delta_{0}}\Big) |\Psi_+\rangle\langle\Psi_+| + \Delta_{0} |\Psi_-\rangle\langle\Psi_-| \nonumber \\
&=& \frac{1}{2}\Big(\Delta_{0} + \frac{\tilde{\xi}^2_{+}}{\Delta_{0}}\Big) (|\Psi_L\rangle + |\Psi_R\rangle)(\langle\Psi_L| + \langle\Psi_R|) +  \frac{\Delta_0}{2}(|\Psi_L\rangle - |\Psi_R\rangle)(\langle\Psi_L| - \langle\Psi_R|) \nonumber \\
&=& \Big(\Delta_0+\frac{\tilde{\xi}^2_{+}}{\Delta_{0}}\Big) (|\Psi_L\rangle\langle\Psi_L| + |\Psi_R\rangle\langle\Psi_R|) + \frac{1}{2}\frac{\tilde{\xi}^2_{+}}{\Delta_{0}}(|\Psi_L\rangle\langle\Psi_R| + |\Psi_R\rangle\langle\Psi_L|).
\end{eqnarray}
\begin{figure}[b]
\includegraphics[width=14cm]{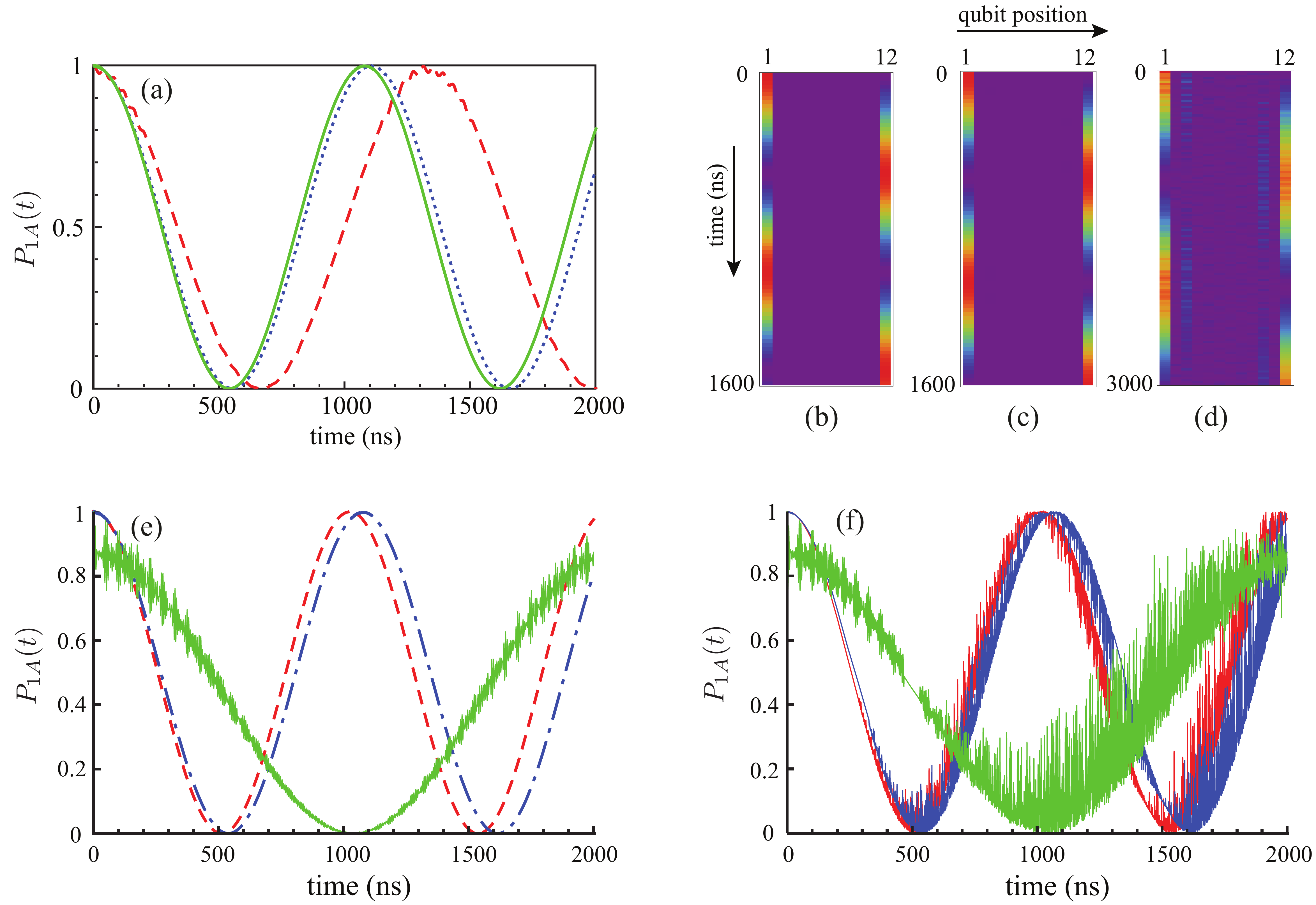}
\caption{(a) Excitation dynamics of the left-edge qubit with $\varphi=0.1\pi$. The curves with red-dashed, blue-dotted and green-solid curves correspond to $t_0/2\pi= 10, 50$ and $100$ MHz, respectively. (b)-(d) Excitation dynamics  in the cavity-coupled SSH qubit array for $\varphi=0.01\pi,0.1\pi$ and $0.3\pi$, respectively. (e) Excitation dynamics of the left-edge qubit for $\varphi=0.01\pi$ (red-dashed), $0.1\pi$ (blue-dash-dotted) and $0.3\pi$ (green-solid). (f) The effect of disorder. Here we consider the disorder strength $\epsilon=0.2\times 2\pi$ MHz. Other parameters are as the same as Fig.~4 in the main text. }\label{dynamics2}
\end{figure}
In this scenario, the coupling between left and right edge states is
\begin{eqnarray}
J'&=&\frac{\tilde{\xi}_+^2}{2\Delta_0} \nonumber \\
&=&\cos\varphi \frac{g_0^2}{\Delta_0}. \label{eqec2}
\end{eqnarray}
In other words, no matter the edge states are hybridized or not, the cavity-mediated couplings between left and right edge states is $\cos\varphi g_0^2/\Delta_0$, as long as the energy splitting between hybridized edge states is negligible comparing to $\Delta_0$. Due to this fact, the cavity can induce nonlocal edge-state couplings in a large range of $\varphi$.

The coupling between edge states provides a way to control topological modes. Topology-protected quantum state transfer can be realized. The Rabi oscillation of the excitation in an edge qubit can be regarded as a signature of the topological coupling. If the initial state is $|\psi_0\rangle=|100\cdots\rangle$, then the fidelity for the revival of the excitation is
\begin{eqnarray}
P_{1A}(t)&=& |\langle\psi_0|e^{-\frac{i}{\hbar}\bar{H} t}|\psi_0\rangle|^2.
\end{eqnarray}

As shown in Fig.~\ref{dynamics2}(a), the excitation dynamics of the qubit at left edge is demonstrated. The periodic oscillation of the excitation is produced by the coupling between edge states. From the oscillation period, we can  obtain the coupling strength. However, we find that the bandgap has influence to the coupling of edge states. When the bandgap is large enough, the edge states have coupling $J$. However, when the bandgap is decreased, the coupling becomes smaller than $J$. In Fig.~\ref{dynamics2}(a), we present the Rabi dynamics for different values of $t_0$, which determines the topological bandgap. Even the bandgap is small, e.g., $t_0/2\pi=10$ MHz, the revival fidelity is high. The reason is that the edge states have large effective coupling, comparing to the couplings between edge and bulk states, which are detuned. Therefore, the edge states form a subspace where the excitation can be exchanged. In Figs.~\ref{dynamics2}(b)-\ref{dynamics2}(d), we show the excitation dynamics for the cavity-coupled qubit array in the topological phase. For $\varphi=0.01\pi$, the edge states are unhybridized in the SSH qubit array (with $6$ unit cells). However, the edge states are hybridized when $\varphi=0.1\pi$ and $0.3\pi$. The excitation dynamics of the left-edge qubit is shown in Fig.~\ref{dynamics2}(e). The cavity induced edge-state coupling $J=\cos\varphi g_0^2/\Delta_0$ yields periodic oscillation of edge qubits. The difference between the unhybridized case, e.g., $\varphi=0.01\pi$, and hybridized case, e.g., $\varphi=0.1\pi$, is produced by the $\varphi$-dependent edge-state coupling (see Eq.~(\ref{eqec1}) and Eq.~(\ref{eqec2})). In Fig.~\ref{dynamics2}(f), we consider the disorder effect of qubits' frequencies on the excitation dynamics. In the main text, when we discuss the cavity-mediated qubit-qubit interactions, the frequency of the cavity is assumed in the topological bandgap. This requirement limits the strengths of the cavity-mediated qubit-qubit interactions. Therefore, the disorder induces fluctuations in the excitation dynamics.

\begin{figure}[!htbp]
\includegraphics[width=14cm]{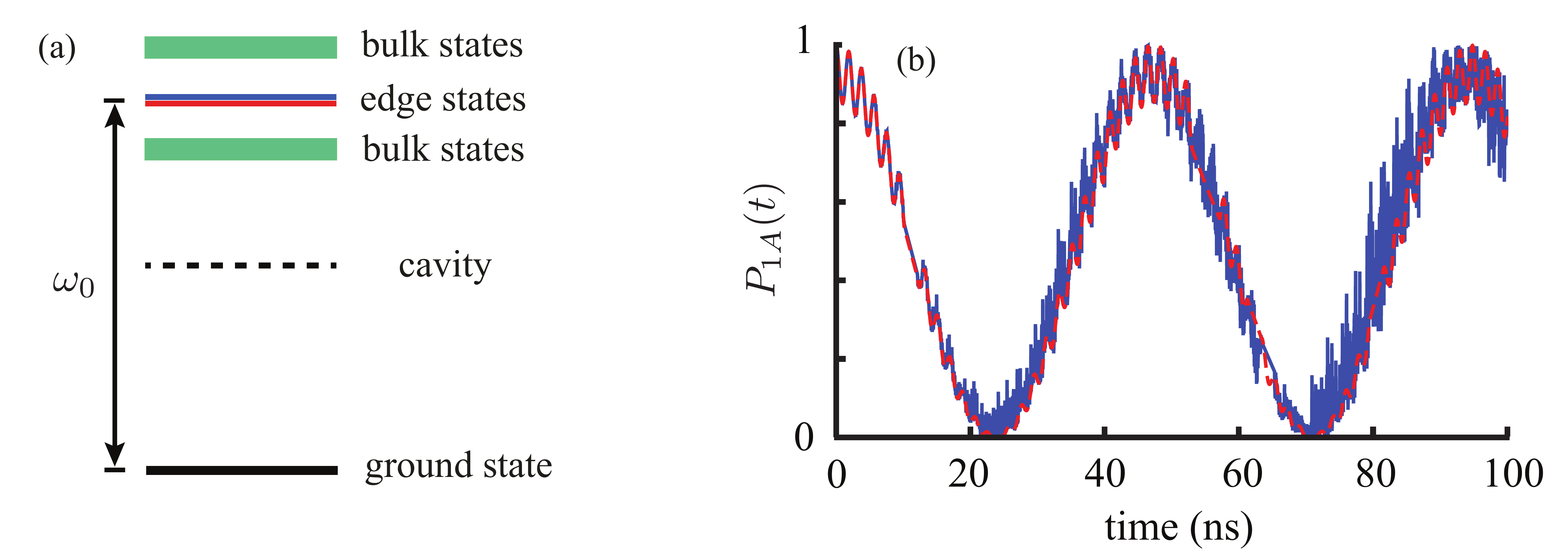}
\caption{(a) Level structure of the qubit array. (b) Excitation dynamics of the left-edge qubit with $g_0=300\times 2\pi$ MHz and $\Delta_0=3\times 2\pi$ GH. Red-dashed and blue-solid lines correspond to the disorder strength $\epsilon=0$ and $\epsilon=4\times 2\pi$ MHz, respectively. Here we consider $N=6, \varphi=0.1\pi$.}\label{figdisorder2}
\end{figure}

In Fig.~\ref{figdisorder2}(a), the frequency of cavity is far away from the topological bandgap. In this case, the cavity can have large gaps with edge states. Because of the topological bandgap, the edge states are not coupled to bulk states. The scheme in Fig.~\ref{figdisorder2}(a) allows us to study strong couplings between cavity and qubits. For example, when $g_0/2\pi=300$ MHz and $\Delta_0/2\pi=3$ GHz, the cavity-mediated qubit-qubit interactions are $g_0^2/\Delta_0=30\times2\pi$ MHz. Accordingly, the large edge-state coupling can be obtained. In Fig.~\ref{figdisorder2}(b), we show the robustness of the excitation dynamics of the left-edge qubit to disorder.

In practical systems, the observation of excitation dynamics is limited by the lifetime of qubits. In particular, the dissipative cavity can lead to decays of qubits, i.e., the Purcell effect. The decay of qubit induced by dissipative cavity is $\kappa g_0^2/\Delta_0^2$. In Figs.~\ref{dynamics2}(b)-\ref{dynamics2}(f), we consider $g_0/\Delta_0=0.1$. Therefore, the qubit decay induced by dissipative cavity is $0.01\kappa$. Therefore, the lifetime of qubits is $1/(\gamma+0.01\kappa)\approx1.33$ $\mu$s. The coupling between edge states is $J=\cos\varphi g_0^2/\Delta_0$. As $\varphi=0.1\pi$, we have $J=0.48\times 2\pi$ MHz and the period of the excitation dynamics if $T=\pi/J\approx1.04$ $\mu$s. Therefore, the edge-state coupling $J$ can be obtained for the lifetime of qubits $1.33$ $\mu$s by measuring the qubit population at the right edge. (The population of qubit at the right edge becomes maximal for half period, i.e., 0.52 $\mu$s). As $\varphi$ increases, the coupling between edge states becomes small. To observe the excitation dynamics for weak edge-state coupling, one could require longer lifetime of qubits, which can be realized by considering cavity with low decay rate.

\subsection{Transparency induced by coupling between edge states}
\begin{figure}[b]
\includegraphics[width=14cm]{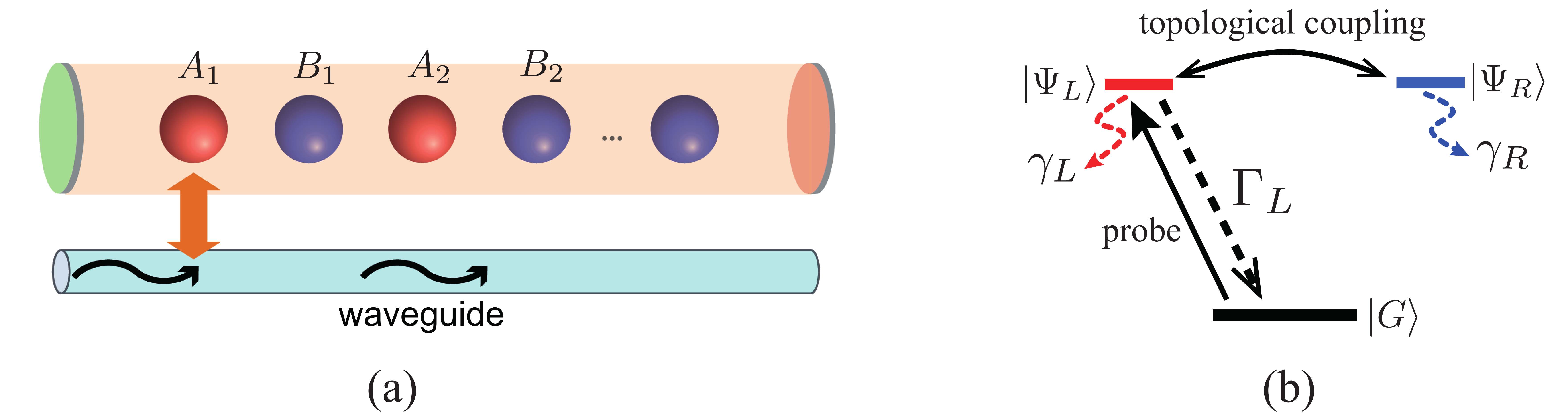}
\caption{(a) Schematic diagram of the waveguide driving of SSH qubit array in a cavity. The qubit at the left edge is coupled to the waveguide. (b) Level structure of (a). The photons in the waveguide interacting with $A_1$ qubit actually couple to the left edge state. And the cavity induced coupling between edge states controls the transmission of probing photons. $\gamma_L$ and $\gamma_R$ are the decay rates for left and right edge states, respectively. And $\Gamma_L$ is the decay rate for left edge state induced by the waveguide. }\label{superatom}
\end{figure}
\begin{figure}[!htbp]
\includegraphics[width=12cm]{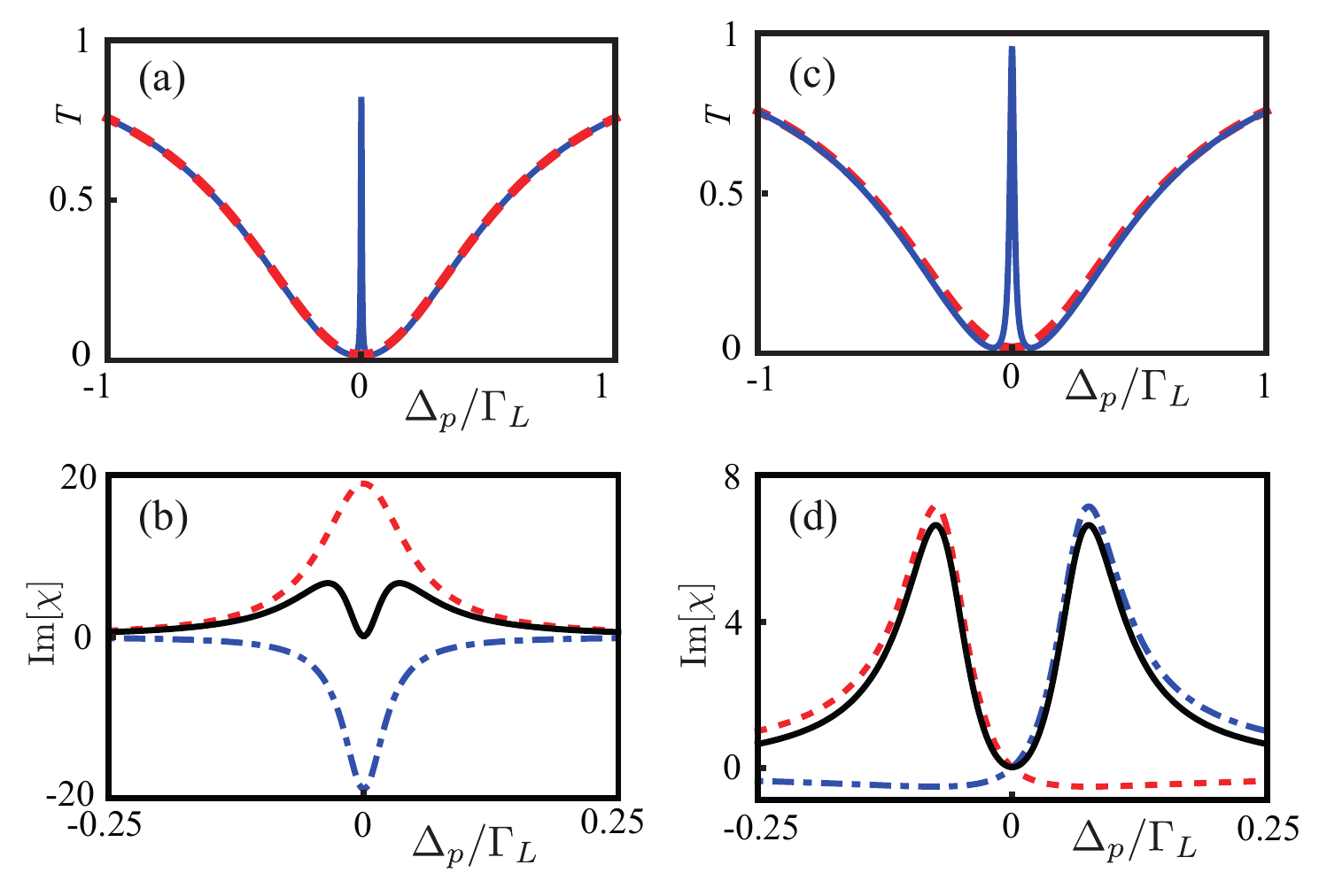}
\caption{(a) The transmission for $J=0$ (red-dahsed) and $J=0.035\Gamma_L$ (blue-solid). (b) The imaginary part of susceptibility (black-solid) for $J=0.035\Gamma_L$, and its decompositions into two peaks denoted by red-dashed and blue-dashed-dotted curves. (c)  The transmission for $J=0$ (red-dahsed) and $J=0.075\Gamma_L$ (blue-solid). (d) The imaginary part of susceptibility (black-solid), and its decompositions into two positive peaks denoted by red-dashed and blue-dashed-dotted curves. Other parameters are $\gamma_L = 0.15 \Gamma_L$, $\gamma_R =5 \times 10^{-4} \Gamma_L$.}\label{susceptibility}
\end{figure}

The nonlocal coupling enables a subspace within two edge states. The edge states are localized to the boundaries of the system, therefore it is feasible to manipulate these edge states. As shown in Fig.~\ref{superatom}(a), we consider a waveguide that couples to $A_1$ qubit at the left edge of the array. When the bulk states are ignored, the SSH qubit array behaves as an effective topological superatom \cite{PhysRevLett.124.023603} with two excited edge state and a ground state. The $A_1$ qubit is protected by the left edge state. Therefore we can find a simple picture to describe the system (see Fig.~\ref{superatom}(b)). Now we present the details of this system. At first, we consider a single mode $\hat{b}$ of the waveguide. The Hamiltonian of the system is
\begin{eqnarray}
H_1/\hbar &=&\omega_0(|\Psi_L\rangle \langle \Psi_L| + |\Psi_R\rangle \langle \Psi_R|) + \omega_c \hat{a}^{\dagger}\hat{a}  +\omega_b \hat{b}^{\dagger}\hat{b} \nonumber \\
&&- g_L (S_{L}^+ \hat{a} + \hat{a}^{\dagger}S_{L}^-) -g_R(S_{R}^+ \hat{a} + \hat{a}^{\dagger}S_{R}^-) - g_b (S_{L}^+ \hat{b} + \hat{b}^{\dagger}S_{L}^-),
\end{eqnarray}
where $S_{L}^+=|\Psi_L\rangle\langle G|$ and $S_{R}^+=|\Psi_R\rangle\langle G|$.  In a rotating frame with $H_{\mathrm{rot}}=\hbar\omega_{c} (\hat{a}^{\dagger}\hat{a} + \hat{b}^{\dagger}\hat{b}  + |\Psi_L\rangle\langle \Psi_L| + |\Psi_R\rangle\langle \Psi_R|)$, the Hamiltonian can be rewritten as
\begin{eqnarray}
H_2/\hbar &=& \Delta_0(|\Psi_L\rangle \langle \Psi_L| + |\Psi_R\rangle \langle \Psi_R|) + (\omega_b-\omega_c) \hat{b}^{\dagger}\hat{b}   \nonumber \\
&&- g_L (S_{L}^+ \hat{a} + \hat{a}^{\dagger}S_{L}^-) -g_R(S_{R}^+ \hat{a} + \hat{a}^{\dagger}S_{R}^-)- g_b (S_{L}^+ \hat{b} + \hat{b}^{\dagger}S_{L}^-).
\end{eqnarray}
When $g_L, g_R \ll \Delta_0$, we can make a Schrieffer-Wolff transformation with
\begin{equation}
U=\mathrm{exp}[M]=\mathrm{exp}\Big[\frac{g_L}{\Delta_0}(\hat{a}^{\dagger}S_{L}^--S_{L}^+\hat{a}) + \frac{g_R}{\Delta_0}(\hat{a}^{\dagger}S_{R}^--S_{R}^+\hat{a})\Big].
\end{equation}
Therefore,
\begin{equation}
H_3=UH_2U^{\dagger}=H_2 + [M, H_2] + \frac{1}{2!}[M, [M, H_2]] + \ldots
\end{equation}
In $[M, H_2]$, there is a special term
\begin{eqnarray}
[M, S_{L}^+\hat{b}+\hat{b}^{\dagger}S_{L}^-] &=& \frac{g_L}{\Delta_0} [\hat{a}^{\dagger}S_{L}^--S_{L}^+\hat{a}, S_{L}^+\hat{b}+\hat{b}^{\dagger}S_{L}^-] + \frac{g_R}{\Delta_0} [\hat{a}^{\dagger}S_{R}^--S_{R}^+\hat{a}, S_{L}^+\hat{b}+\hat{b}^{\dagger}S_{L}^-] \nonumber \\
&=& \frac{g_L}{\Delta_0} (|G\rangle\langle G|-|\Psi_L\rangle \langle \Psi_L|)(\hat{a}^{\dagger} \hat{b} + \hat{b}^{\dagger}\hat{a}) -\frac{g_R}{\Delta_0}(\hat{a}^{\dagger}\hat{b}|\Psi_L\rangle \langle \Psi_R|+ |\Psi_R\rangle \langle \Psi_L| \hat{b}^{\dagger}\hat{a}).
\end{eqnarray}
We assume $\langle \hat{a}^{\dagger} \hat{a} \rangle=0$, i.e., mode $\hat{a}$ is in the vacuum state. Therefore, the above term is zero. So, we can obtain the effective Hamiltonian
\begin{eqnarray}
H_3/\hbar &=& \Big(\Delta_0 + \frac{g_L^2}{\Delta_0}\Big)|\Psi_L\rangle \langle \Psi_L| + \Big(\Delta_0+\frac{g_R^2}{\Delta_0}\Big)|\Psi_R\rangle \langle \Psi_R| + J (|\Psi_L\rangle \langle \Psi_R| + \mathrm{H.c.})\nonumber \\
 && + (\omega_b-\omega_c)\hat{b}^{\dagger}\hat{b} -g_b(S_{L}^+\hat{b}+\hat{b}^{\dagger}S_{L}^-), \label{EIT1}
\end{eqnarray}
where $J=g_L g_R/\Delta_0$ represents the coupling between edge state. When all the modes in the waveguide are considered, we obtain
\begin{eqnarray}
H_4/\hbar &=&\Big(\Delta_0 + \frac{g_L^2}{\Delta_0} - i \frac{\gamma_L}{2}\Big)|\Psi_L\rangle \langle \Psi_L| + \Big(\Delta_0+\frac{g_R^2}{\Delta_0} - i \frac{\gamma_R}{2}\Big)|\Psi_R\rangle \langle \Psi_R| + J(|\Psi_L\rangle \langle \Psi_R| + \mathrm{H.c.})  \nonumber \\
 && + \sum_n (\omega_{b,n}-\omega_c)\hat{b}^{\dagger}_n\hat{b}_n -g_{b} \sum_n (S_{L}^+\hat{b}_n+\hat{b}^{\dagger}_n S_{L}^-), \label{EIT2}
\end{eqnarray}
where $\gamma_L$ and $\gamma_R$ denote the dissipations of left and right edge states produced by parasitic environmental modes. And $\omega_{b,n}$ denote the frequencies of photonic modes in the waveguide, $n$ is the index for different modes. For simplicity, we assume linear dispersion relation of the modes in waveguide, i.e., $\omega_b=c k$. In a 1D waveguide, photons propagate along left or right direction.  So the Hamiltonian of the waveguide is
\begin{equation}
H_{\mathrm{wg}}/\hbar = -i c \int dx \Big(\hat{b}_R^{\dagger}(x) \frac{\partial}{\partial x} \hat{b}_R(x) -  \hat{b}_L^{\dagger}(x) \frac{\partial}{\partial x} \hat{b}_L(x)\Big).
\end{equation}
Considering the symmetric and anti-symmetric superpositions of left and right propagating photonic modes \cite{PhysRevLett.98.153003},
\begin{eqnarray}
\hat{b}_e(x) &=& \frac{1}{\sqrt{2}} (\hat{b}_R(x) + \hat{b}_L(-x)), \\
\hat{b}_o(x) &=& \frac{1}{\sqrt{2}} (\hat{b}_R(x) - \hat{b}_L(-x)),
\end{eqnarray}
the Hamiltonian of the waveguide can be written as
\begin{equation}
H_{\mathrm{wg}}/\hbar = -i c \int dx \hat{b}_e^{\dagger}(x) \frac{\partial}{\partial x} \hat{b}_e(x).
\end{equation}
Therefore, Eq.~(\ref{EIT2}) can be rewritten as
\begin{eqnarray}
H_5/\hbar &=& \Big(\Delta_0 + \frac{g_L^2}{\Delta_0} - i \frac{\gamma_L}{2}\Big)|\Psi_L\rangle \langle \Psi_L| + \Big(\Delta_0 + \frac{g_R^2}{\Delta_0}- i \frac{\gamma_R}{2}\Big)|\Psi_R\rangle \langle \Psi_R| + J (|\Psi_L\rangle\langle \Psi_R| + |\Psi_R\rangle \langle \Psi_L|)\nonumber \\
 &&+\int dx \hat{b}_e^{\dagger}(x)\Big(-i c \frac{\partial}{\partial x} -\omega_c \Big) \hat{b}_e(x) - \tilde{g}_b\int dx \delta(x)[\hat{b}_e^{\dagger}(x)|G\rangle\langle \Psi_L| + \mathrm{H.c.}],
\end{eqnarray}
where $\tilde{g}_b=\sqrt{2}g_b$. The state of the system has the form (in the single-photon subspace)
\begin{equation}
|\Psi(t)\rangle = \int dx f_1(x) \hat{b}_e^{\dagger}(x)|\emptyset,G\rangle + f_2|\emptyset,\Psi_L\rangle + f_3|\emptyset,\Psi_R\rangle.
\end{equation}
Here, $|\emptyset\rangle$ represents empty waveguide without photons. Considering the following ansatz,
\begin{equation}
f_1(x) = \frac{1}{\sqrt{2\pi}} e^{ikx}[\theta(-x)+t_k \theta(x)],
\end{equation}
where $\theta(x)$ is the Heaviside step function. Solving the Schr\"{o}dinger equation, we can obtain
\begin{equation}
t_k = \frac{[i(ck-\tilde{\omega})+\frac{\Gamma_L}{2}-\frac{\gamma_{L}}{2}]  [i(ck-\tilde{\omega})-\frac{\gamma_{R}}{2}] + J^2}{[i(ck-\tilde{\omega})-\frac{\Gamma_{L}}{2}-\frac{\gamma_{L}}{2}]  [i(ck-\tilde{\omega})-\frac{\gamma_{R}}{2}] + J^2},
\end{equation}
with $\tilde{\omega}=\omega_0+g_L^2/\Delta_0$. Here $\Gamma_L$ is the decay rate of left edge state produced by the waveguide. Therefore the transmission amplitude of the probing photon (with frequency $\omega$) is
\begin{eqnarray}
t&=&\frac{t_k+1}{2} \nonumber \\
&=&\frac{[i\Delta_p-\frac{\gamma_{L}}{2}]  [i\Delta_p-\frac{\gamma_{R}}{2}] + J^2}{[i\Delta_p-\frac{\Gamma_{L}}{2}-\frac{\gamma_{L}}{2}]  [i\Delta_p-\frac{\gamma_{R}}{2}] + J^2},
\end{eqnarray}
with
\begin{eqnarray}
\Delta_p&=&ck-\tilde{\omega} \nonumber \\
&=&ck-\omega_0-\frac{g_L^2}{\Delta_0}.
\end{eqnarray}
In the waveguide-driven atoms, transparency of probing photons~\cite{Witthaut2010} and single-photon frequency conversion~\cite{Jia2017} can be realized. The transparency can be produced by two different effects, i.e., quantum interference and energy splitting. Here, we consider how to distinguish them in our setup. Before that, we define the susceptibility
\begin{equation}
\chi=-i(1-\frac{1}{t}).
\end{equation}
In Fig.~\ref{susceptibility}(a), we show the transmission spectrum for $J=0$ and $J=0.035\Gamma_L$, respectively. And the coupling between edge states leads to transparency of probing photons. This transparency is produced by the quantum interference which can be seen from Fig.~\ref{susceptibility}(b). The imaginary part of susceptibility is composed of two peaks, one is positive, the other is negative. This is the feature from quantum interference~\cite{PhysRevA.81.041803,PhysRevA.89.063822}. In Fig.~\ref{susceptibility}(c), we consider $J=0$ and $J=0.075\Gamma_L$. We also find a transparency when the coupling between edge states is considered. However, this transparency is not a quantum interference effect. As shown in Fig.~\ref{susceptibility}(d), the imaginary part of the susceptibility can be decomposed into two positive peaks, which manifest the energy splitting produced by the edge-state coupling.

\end{document}